\documentclass[a4paper,fleqn,usenatbib]{mnras}
\usepackage{newtxtext,newtxmath}
\usepackage[T1]{fontenc}
\usepackage{ae,aecompl}

\usepackage{diagbox}
\usepackage{pbox}
\usepackage{multirow}  

\usepackage[hang,raggedright,tight]{subfigure}
\usepackage{longtable}
\usepackage[figuresright]{rotating}
\usepackage{float}
\usepackage[toc,page]{appendix}

\usepackage{subfigure}
\usepackage{mathastext}
\usepackage{gensymb}

\title[Measuring ISW effect with LDPs]{Measuring the integrated Sachs-Wolfe effect from the low-density regions of the universe}

\author[Dong et al.]{
Fuyu Dong,$^{1,2,4}$
Yu Yu,$^{1,2}$
Jun Zhang,$^{1,2}$\thanks{betajzhang@sjtu.edu.cn}
Xiaohu Yang,$^{1,2,3}$
Pengjie Zhang,$^{1,2,3}$\thanks{zhangpj@sjtu.edu.cn}
\\
$^{1}$Department of Astronomy, School of Physics and Astronomy, Shanghai Jiao Tong University, Shanghai, 200240,China\\
$^{2}$Shanghai Key Laboratory for Particle Physics and Cosmology, Shanghai, 200240, China\\
$^{3}$Division of Astronomy and Astrophysics, Tsung-Dao Lee Institute, Shanghai Jiao Tong University, Shanghai, 200240, China\\
$^{4}$School of Physics, Korea Institute for Advanced Study (KIAS), 85 Hoegiro, Dongdaemun-gu, Seoul, 02455, Republic of Korea
}

\date{Accepted XXX. Received YYY; in original form ZZZ}
\pubyear{2020}
\begin{document}
\label{firstpage}
\pagerange{\pageref{firstpage}--\pageref{lastpage}}
\maketitle

\begin{abstract}
The integrated Sachs-Wolfe (ISW) effect is caused by the decay of cosmological gravitational potential, and is therefore a unique probe of dark energy. However, its robust detection is still problematic. Various tensions between different data sets, different large scale structure (LSS) tracers, and between data and the $\Lambda$CDM theory prediction, exist. We propose a novel method of ISW measurement by cross correlating CMB and the LSS traced by ``low-density-position'' (LDP, \citet{2019ApJ...874....7D}).  It isolates the ISW effect generated by low-density regions of the universe, but insensitive to selection effects associated with voids. 
We apply it to the DR8 galaxy catalogue of the DESI Legacy imaging surveys, and obtain the LDPs at $z\leq 0.6$ over $\sim$ 20000 $deg^2$ sky coverage. We then cross correlate with the Planck temperature map, and detect the ISW effect at $3.2\sigma$.  We further compare the measurement with numerical simulations of the concordance $\Lambda$CDM cosmology, and find the ISW amplitude parameter $A_{ISW}=1.14\pm0.38$ when we adopt a LDP definition radius $R_s=3^{'}$, fully consistent with  the  prediction of the standard $\Lambda$CDM cosmology ($A_{ISW}=1$).  This agreement with $\Lambda$CDM cosmology holds for all the galaxy samples and $R_s$ that we have investigated. Furthermore, the S/N is comparable to that of galaxy ISW measurement. These results demonstrate  the LDP method as a competitive alternative to existing ISW measurement methods, and provide independent checks to  existing tensions.
\end{abstract}


\section{INTRODUCTION}
\label{introduction}
The integrated Sachs-Wolfe (ISW) effect \citep{1967ApJ...147...73S} probes the time variation of gravitational potential, through the induced CMB temperature fluctuation
\begin{equation}
\label{dT2}
\Delta T(\widehat{n})=\frac{2}{c^3}T_0\int\dot{\Phi}(r,\widehat{n})\,a\,dr \ .
\end{equation}
Here $\widehat{n}$ is the line of sight,  $T_0$ the mean temperature of CMB, $c$  the speed of light, $\dot{\Phi}$ the time derivative of the gravitational potential along, $a$ the cosmic scale factor and $r$ the comoving radial distance.  The gravitational potential $\Phi$ at large/linear scale is time independent, if gravity is GR, the universe is flat and the total matter density $\Omega_m=1$. Observations of primary CMB \citep{2016A&A...594A..13P} show that our universe is flat. Then within the framework of general relativity,  any detection of the ISW effect would serve as a smoking gun of dark energy.  It can then be used to constrain the dark energy equation of state, and even clustering of dark energy around horizon scale (e.g. \cite{2003MNRAS.346..987W,2004PhRvD..69h3503B,2004PhRvD..70l3002H,10.1111/j.1745-3933.2006.00218.x,2008ApJ...675...29M,2010PhRvD..81j3513D}). Alternatively, it can be used to test GR at cosmological scales \citep{2003AnPhy.303..203H,2006PhRvD..73l3504Z,2006PhRvD..74f3520G,2007MNRAS.381.1347C,2008PhRvD..78h7303F,2012MNRAS.426.2581G}, constrain primordial non-Gaussianities \citep{2012JCAP...06..042N}, or probe GR backreaction \citep{2017MNRAS.469L...1R}.

The major factor limiting the cosmological application of ISW is its weak signal, overwhelmed by the primary CMB. It can be separated from primary CMB, by cross-correlating  with the large scale structure (LSS) \citep{1996PhRvL..76..575C,2000ApJ...538...57S}. However, to further suppress the cosmic variance, CMB surveys of nearly full sky coverage and wide and deep galaxy surveys are both required. From the release of the first year WMAP data, there have been various works to measure the ISW effect \citep{2003ApJ...597L..89F,2004Natur.427...45B,2004PhRvD..70h3536A,2004PhRvD..69h3524A,2004MNRAS.350L..37F,2004ApJ...608...10N,2005NewAR..49...75B,2005PhRvD..72d3525P,2005PhRvD..71l3521C,2006MNRAS.365..891V,2006astro.ph..2398M,2007MNRAS.381.1347C,2007MNRAS.377.1085R,2008MNRAS.386.2161R,2010A&A...520A.101H,2010MNRAS.404..532M,2012MNRAS.427.3044S,2012MNRAS.426.2581G,2014A&A...571A..19P,Shajib_2016,2016A&A...594A..21P}. The analyzed CMB experiments include both WMAP and Planck. The analyzed galaxy surveys include SDSS, NVSS, 2MASS, WISE, etc. The LSS tracers include SDSS main galaxies and luminous red galaxies in optical bands, radio galaxies, AGNs, and even weak gravitational lensing reconstructed from CMB \citep{2016A&A...594A..21P}. Although some claimed detection significances ($\sim 2$-$4\sigma$) may be questionable\citep{2012MNRAS.426.2581G,2014MNRAS.438.1724H}, these measurements are in general  consistent with the $\Lambda$CDM prediction.

Besides directly using galaxies as LSS tracers, entities derived with galaxy surveys such as superclusters and voids are also explored to measure the ISW effect \citep{2008ApJ...683L..99G,2011ApJ...732...27P,2014A&A...571A..19P}. These measurements have different S/N and different systematics, and therefore are highly complementary to ISW measurements with galaxies. For example, ISW measured from voids may have less contamiantion from radio emission associated with galaxies. However, there are tensions betwen existing measurements and between measurements and theoretical prediction. For example, \cite{2008ApJ...683L..99G} stacked the most significant 50 clusters/voids (superclusters/supervoids) identified with SDSS photo-z data, and found a $4\sigma$ detction of the ISW effect.  However, after investigated by other papers, this detection has been found difficult to explain \citep{2010MNRAS.401..547H,2010ApJ...724...12I,2014A&A...572C...2I,2015MNRAS.446.1321H,2015MNRAS.452.1295K}.

Tensions persists in later measurements with different void catalogues. In most cases excess signals are reported compared to the $\Lambda CDM$ prediction \citep{2012JCAP...06..042N}. For example, \cite{2014ApJ...786..110C} found that their detection with SDSS voids is at odds with simulations of a $\Lambda CDM$ universe at $\sim 2\sigma$. In term of the ISW amplitude $A_{ISW}$ ($A_{ISW}=1$ in the standard $\Lambda$CDM cosmology),  \citet{2018MNRAS.475.1777K} found  with BOSS supervoids $A_{ISW} \approx 9$. \citet{2019MNRAS.484.5267K} found with DES super-voids $A_{ISW} \approx 4.1$.

These tensions are unlikely caused by new physics beyond $\Lambda$CDM, since the ISW measurements with galaxies in the same surveys/cosmic volumes are usually consistent with the $\Lambda$CDM prediction. Furthermore, it is found that the ISW measurements with voids rely heavily on the void catalogues (e.g. \citet{2008ApJ...683L..99G, 2015MNRAS.452.1295K}) and therefore the associated systematics, if uncorrected. In addition,  null detections were also reported in observations and favored in simulations \citep{2015MNRAS.446.1321H,2014A&A...572C...2I}. 

Among all possibilities leading to the above controversies, void identification in observations likely plays a major role. Voids are defined as low number density regions in the galaxy distribution field. But in low density regions, the galaxy number distribution suffers from relatively larger shot noise. This makes the identification of voids and their centers/radii difficult. The low number density further amplifies the impact of  non-uniform galaxy (radial and angular) selection function in spectroscopic redshift surveys, and making its correction more challenging than the case of galaxy clustering. Situation becomes even worse for the photometric data, for which the smearing effect of redshift errors of galaxies in the line-of-sight will ``merge" voids. These issues also complicate the correspondence between voids in observations and in simulations/theory, and make the theoretical interpretation difficult.

Taking these issues into account,  in this paper we revisit the measurement of ISW effect by considering a new LSS tracer -- ``low-density-position'' (LDP)  \citep{2019ApJ...874....7D}. LDPs are the collection of sky positions, after removing positions within a given radius of any observed galaxy. Statistically speaking, they correspond to low density regions. The density threshold depends on the radius to perform the cut and the mean number density of observed galaxies. So by varying the cut radius, LDPs can also serve as intermediate case between galaxies and voids. So they will provide independent check on the above tensions found between voids and the concordance $\Lambda$CDM, and between observations of voids and galaxies. Furthermore,  comparing to voids,  LDPs are more straightforward to identify in both observations and in simulations, making the data interpretation more reliable. In \cite{2019ApJ...874....7D}, we have used LDPs to achieve significant detection of weak lensing using CFHTLenS data, and to differentiate dark energy models. In this work, we treat LDPs as density tracers and focus on the  LDP-temperature correlation measurement. Our aim is to provide independent ISW measurement, which is then used to test the concordance $\Lambda$CDM cosmology, and cross-check with existing ISW measurements with voids. 

The paper is organized as follows: In \S\ref{sec:obs-data}, we introduce the data sets and procedures to measure the LDP-temperature correlations ($\omega_{T\ell}$). In \S\ref{sec:simu-data}, we calculate the theoretical prediction from mocks generated by a $\Lambda CDM$ N-body simulation. \S$\ref{sec:result}$ shows our main results of ISW measurement with the LDP method. In \S\ref{sec:conclusion}, we give conclusion and discussions about related issues. 
Some further technical details are presented in the appendix, along with the measured CMB-galaxy correlation (appendix\ref{appendix:ISW-galaxy}).

\section{Operating with the Observational Data}
\label{sec:obs-data}
We choose the DR8 galaxy catalogue of the DESI imaging surveys to construct the LDP field, and then cross with Planck SMICA map to measure the ISW effect in the low-density regions of the universe. 
\subsection{Galaxy Catalogue and LDP generation}
\label{sec:obs}
\begin{figure}
    \centering
    \subfigure{
     \includegraphics[width=1\linewidth, clip]{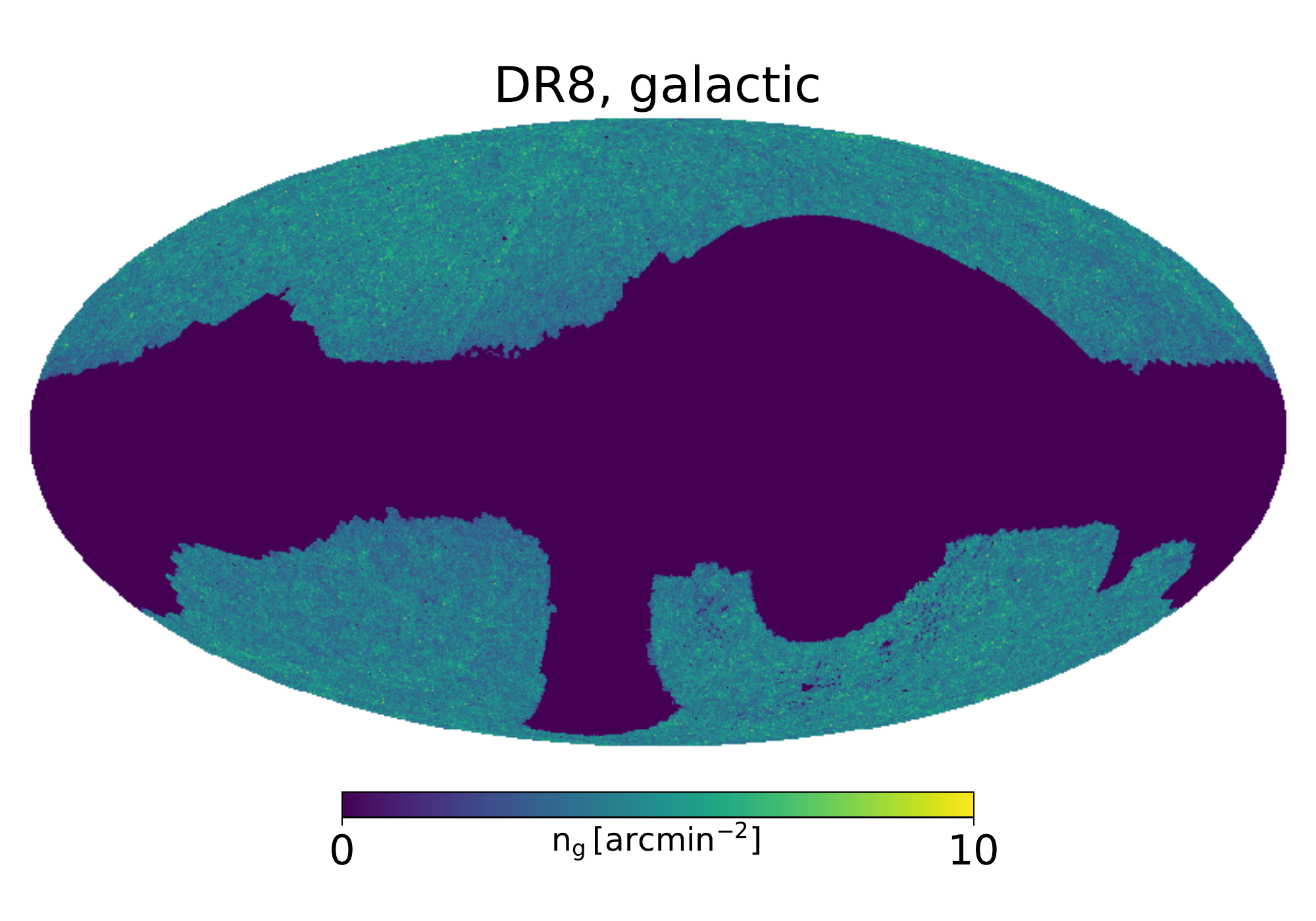}}
   \caption{Source distribution in the DR8 catalogue of the BASS + MzLS + DECaLS + DES imaging surveys. The depth of color represents the number of galaxies per $arcmin^2$.}
    \label{fig:DR8}
\end{figure}
The DR8 galaxy catalogue\footnote{\url{http://batc.bao.ac.cn/~zouhu/doku.php?id=projects:desi_photoz:;}} is a combination of four surveys: BASS\citep{2019ApJS..245....4Z}, MzLS\citep{2016AAS...22831702S}, DECaLS\citep{2005astro.ph.10346T,2016AAS...22831701B} and DES\citep{2018ApJS..239...18A}. They are independent optical imaging surveys with close photometric systems. The first three  together with the infrared WISE survey \citep{2010AJ....140.1868W} aim at providing galaxy and quasar targets for the follow-up Dark Energy Spectroscopic Instrument survey (DESI; \cite{2016arXiv161100036D}). In the DR8 data release, BASS+MzLS locates in the north Galactic cap, DES locates in the south Galactic cap, and DECaLS locates in both north and south Galactic caps along the equator, resulting in a joint sky coverage $\sim 20000 \;deg^2$. The galaxy catalogue provides the photometrically estimated redshifts (hereafter photo-z), apparent magnitudes in g,r,z bands and stellar masses of galaxies. It is the largest galaxy data set  currently available. Its large sky coverage is useful to reduce the statistical errors in our measurements,  a key to improve the ISW measurement. 

The galaxy catalogue includes those sources which have detections in g, r and z bands, and $r<23$. Furthermore,  stars have been excluded through star-galaxy classification.  As we can see from Fig.\ref{fig:DR8},  the surface density of galaxies is reasonably uniform across the sky. The residual non-uniformity (inhomogeneous selection function) is a severe issue for measuring the galaxy auto-correlation. But it is less an issue for cross-correlation measurements presented here, since the selection function is largely uncorrelated with the LSS and its impact on ISW-galaxy cross correlation is taken into account in the random catalogue and in the simulation mocks. 
Coordinates of galaxies have been converted from equatorial coordinate system into galactic coordinate system, in which all the operations are done in the rest of the paper. 

\begin{figure}
    \centering
    \subfigure{
     \includegraphics[width=1\linewidth, clip]{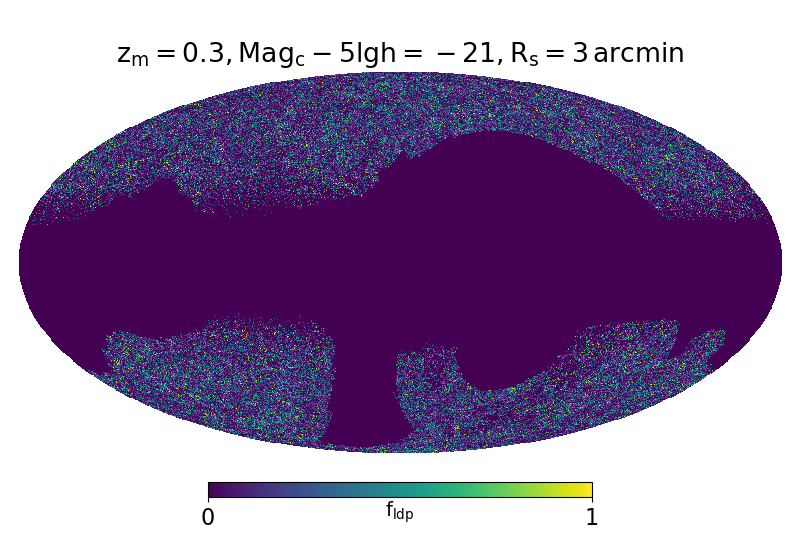}}
   \caption{The full-sky $f_{ldp}$ (the fraction of low-density points in each cell/pixel) distribution. LDPs are generated with cut radius $R_s=3'$, for the galaxy sample with magnitude cut $Mag_c-5lgh=-21$ and photo-z cut $0.2<z<0.4$. }
    \label{fig:ldp}
\end{figure}

\begin{table*}[!htp]
    \footnotesize
    \centering
    \caption{The galaxy samples we use in the DR8 catalogue.}
    \label{table:DR8}
\begin{tabular}{cccccccccccccccc}
\hline
& \multicolumn{4}{c}{$n_{gal}\,(10^6)$ }&&$R_s$ [arcmin] &\multicolumn{4}{c}{$\overline{f}_{ldp}$}&&\multicolumn{4}{c}{$\delta_{l,max}$}\\  
\hline
 {\diagbox[innerleftsep=-0.5cm,innerrightsep=0pt,innerwidth=3.6cm]{$z_m$}{$Mag_c-5lgh$}} &  \multicolumn{1}{c}{-20.5} &-21 &-21.5 & -22 &&& -20.5 &-21 &-21.5 & -22 && -20.5 &-21 &-21.5 & -22\\
\hline
\multirow{2}{*}{0.1} &\multirow{2}{*}{1.73}   &\multirow{2}{*}{0.81}   &\multirow{2}{*}{0.28}      &\multirow{2}{*}{-}       &&3& 0.55  & 0.75   & 0.9     &-         &&  0.81    &0.34 &0.11   &- \\
&&&&&&5&  0.22 &0.47 &0.75 &- &&3.44 &1.13 &0.33 &-\\
\hline
\multirow{2}{*}{0.3} &\multirow{2}{*}{-}       &\multirow{2}{*}{4.53}   &\multirow{2}{*}{1.31}       & \multirow{2}{*}{0.26} &&3& -        & 0.22  &0.62     & 0.9     &&  -          &3.56 &0.62   &0.11\\
&&&&&&5& - &0.02 &0.28 &0.75 &&- &42.7 &2.51 &0.32 \\
\hline
\multirow{2}{*}{0.5} &\multirow{2}{*}{-}      &\multirow{2}{*}{-}         &\multirow{2}{*}{1.66}       &\multirow{2}{*}{0.24}   &&3& -        & -        & 0.55   & 0.91   && -           &-       &0.83   &0.1\\ 
&&&&&&5&- &- &0.21 &0.77  &&- &- &3.86 &0.3\\
\hline
\multicolumn{14}{l}{`{$n_{gal}$}' is the number of galaxies in each galaxy sample.}&\\
\multicolumn{8}{l}{`{$\overline{f}_{ldp}$}' is the average value of $f_{ldp}$.}&\\
\multicolumn{8}{l}{$\delta_{l,max}$ is the maximum $\delta_l$.}&\\
\end{tabular}
\end{table*}

\subsubsection{LDP identification}
\label{sec:LDP-def}
We identify LDPs and define the associated LSS field through the following procedures:\\

\noindent \emph{\small $\bullet$ Generating Survey Masks}.
Sets of uniformly distributed random catalogues are provided in the DR8 website\footnote{\url{http://legacysurvey.org/dr8/files/\#random-catalogs}}. Each random point contains the exposure times for g, r, z bands based upon the sky coordinate drawn independently from the observed distribution. We choose random points whose exposure times in all three bands are greater than zero and MASKBITS\footnote{MASKBITS greater than zero means that the source at this position overlaps with bad or saturated pixels, like bright star mask, globular cluster and so on.} equals to zero to produce the survey masks which populate the same sky coverage and geometry with the galaxy catalogue. \\

\noindent \emph{\small $\bullet$ Generating LDPs}. LDPs depend on the galaxy sample, so we need to first select the galaxy sample for LDP generation. (1) First, we calculate the r-band absolute magnitudes of galaxies with their apparent magnitudes and photo-z.\footnote{The absolute magnitudes used have not been K-corrected. Since galaxies within the same photo-z bin have similar K-correction and the LDP generation is only sensitive to relative brightness between these galaxies, this lack of K-correction is not an issue for our purpose. }   (2) Then similar to the approach in \cite{2019ApJ...874....7D}, we select galaxies brighter than a certain absolute magnitude, and within a photo-z band $[z_m-0.1,z_m+0.1]$ to form the galaxy sample for LDP generation. $\Delta z=0.2$ is chosen as the photo-z error dispersion  $\lesssim 0.1$ on average. (3) Within this galaxy sample, we circle around each galaxy with an angular radius $R_s$, and remove all positions within this radius from the sky. The remaining regions are defined as LDPs candidates. We also remove the LDP candidates lying within the masks. There are no galaxies within radius $R_s$ to any given LDPs, otherwise these points will be excluded by the LDP definition procedure.  The underlying density in this region, with $\bar{N}$ galaxy on the average over the survey volume but $N=0$ galaxy for the selected LDPs,  has a PDF $P(\delta|N=0)\propto \exp(-\bar{N}(1+\delta))P(\delta)$. $P(\delta)$ peaks at $\delta_{peak}<0$. $P(\delta|N=0)$ then peaks at $\delta<\delta_{peak}<0$. Therefore indeed LDPs occupy under-dense regions. (4) We put LDPs on uniform grids generated with HEALPix\citep{2005ApJ...622..759G}.  Although the finer the grids, the more accurate the LDP distribution can be obtained, we adopt $N_{side}=4096$ HEALPix resolution due to the limitation of the number of random points.

\subsubsection{The LDP over-density maps}
Different LDPs correspond to different under-dense regions. For a given LDP, the distance to the nearest galaxy  $R_{min}$ satisfies $R_{min}\geq R_s$. Statistically speaking, the larger the $R_{min}$, the more negative the underlying matter overdensity $\delta_m$.  Therefore to improve the S/N of ISW-LDP measurement, we need to put larger weight for LDPs with larger $R_{min}$.  Theoretically speaking there exists an optimal mapping between $R_{min}$ and $\delta_m$,  and we should find and use that relation to figure out the optimal weighting.  We leave this issue for futher investigation. Here we take a suboptimal, nevertheless workable,  weighting scheme.

We smooth the LDPs with coarse grids,  and define the LDP over-density field in each cell as:  
\begin{equation}
\label{eqn:LDPoverdensity}
\delta_{l}=\frac{f_{ldp}-\overline{f}_{ldp}}{\overline{f}_{ldp}},
\end{equation}
Here we divide the whole sky into lower resolution cells with $N_{side}=512$ (comparing to $N_{side}=4096$ previously). The corresponding cell size is 6.87'.  $f_{ldp}$ is the area proportion of LDPs occupying the given cell.  It equals  the number of LDPs in each cell devided by $n_{grid}$. $n_{grid}$ is the number of fine grids within the coarse cell, which equals 64 here.  The maximum $\delta_l$ occurs for those cells completely occupied by LDPs, corresponding to regions with the most negative $\delta_m$. In order to reduce the impact of survey mask and edge effect on our calculation, we make a selection on the cells being used. We require that the random points which do not satisfy the criteria in \S\ref{sec:LDP-def} should take up less than $\eta=30\%$ of the area of cells. Otherwise, the cells will be disregarded.  Smaller $\eta$ results into less cells used for the measurement and therefore larger statistical errors. Larger $\eta$ results into larger misidentification of low-density regions and therefore weaker ISW signal.  The adopted $\eta=30\%$ is a balance between the two. As tested in simulation, with this ratio the magnitude of our measured LDP-ISW signal being depressed is less than 4\% compared to the one without mask effect. In observation, we also find that this way of cell-selection helps to enhance the amplitude of $\omega^o_{Tl}$.

In this procedure, the fewer galaxies are selected,  the more LDPs would be produced, and vice versa. Taking this into consideration, we control the size of galaxy samples so that their distribution is neither too crowded nor too sparse, otherwise the $\delta_{LDP}$ generation will lack of accuracy in statistics. For $z_m=0.1$, we consider setting $Mag_c-5lgh$ respectively to be -20.5, -21 and -21.5. For $z_m=0.3$, $Mag_c-5lgh$ is set to -21, -21.5 and -22. For $z_m=0.5$, $Mag_c-5lgh$ is set to -21.5 and -22. Galaxies with redshift less than 0.01 are not used due to their too high number density. In Table \ref{table:DR8} we introduce these galaxy samples and LDPs generated with them. Fig.\ref{fig:ldp} shows one example of the $f_{ldp}$ distribution, for which LDPs are generated with parameters $z_m=0.3$, $Mag_c-5lgh=-21$ and $R_s=3'$.

\subsection{CMB  Data}
We use the Planck  SMICA map for the ISW-LDP measurement. The SMICA map has removed secondary CMB anisotropies such as the thermal Sunyaev Zel'dovich effect, which is also correlated with LSS. It also removes galactic foregrounds, which may be correlated with the galaxy mask/selection and therefore may bias the cross correlation measurement. We downgrade the map from $N_{side}=2048$ to $N_{side}=512$ resolution and adopt WMAP 9-year CMB mask \citep{2013ApJS..208...19H,2019MNRAS.484.5267K} to mask out pixels contaminated by Galactic dust or known points
sources. We find that this resolution downgrading only influences the cross-correlation signals at angular separations less than 10 armcin, as we would expect from the large scale origin of the ISW effect.

\begin{figure}
    \centering
    \subfigure{
     \includegraphics[width=1\linewidth, clip]{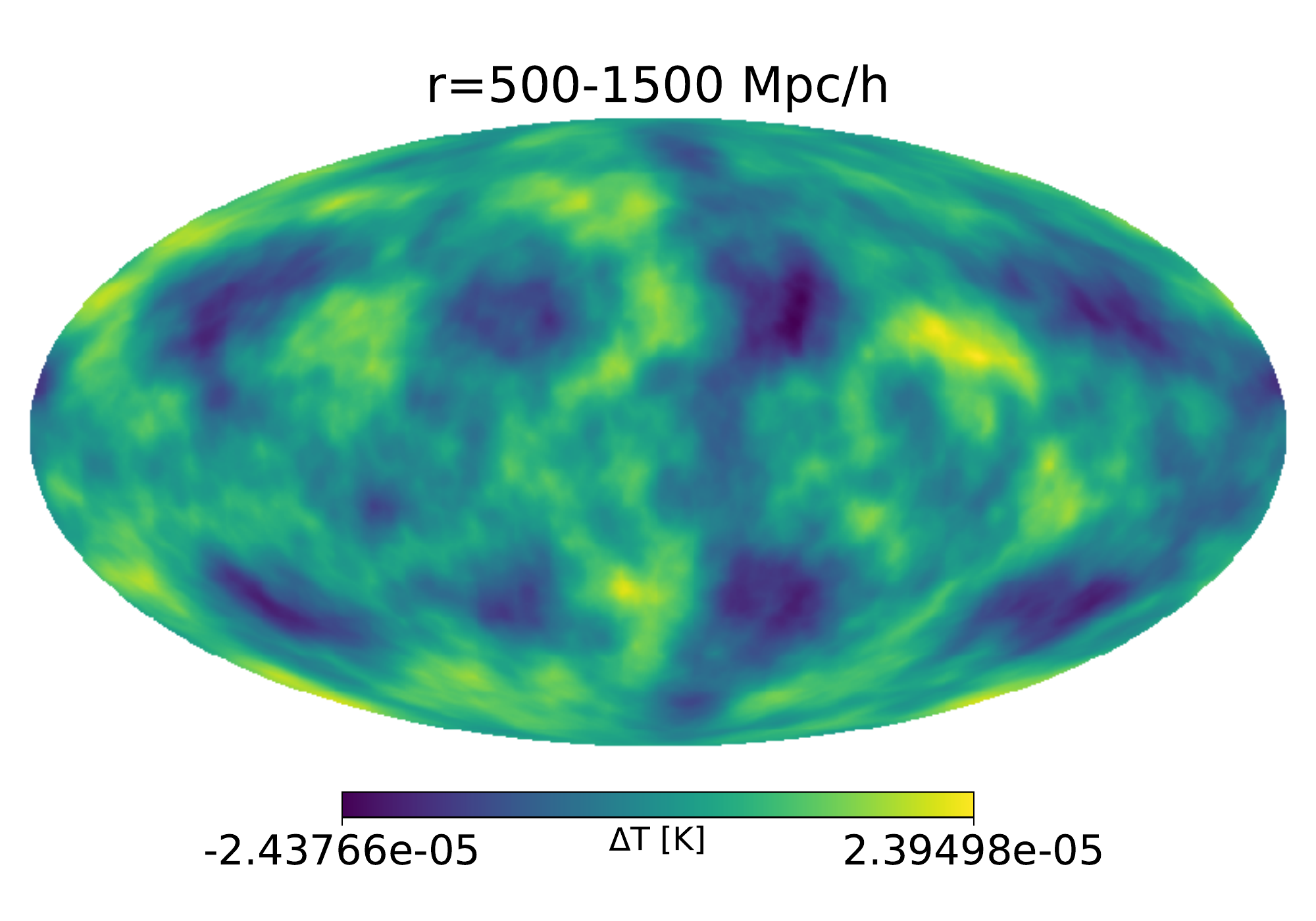}}
       \subfigure{
     \includegraphics[width=1\linewidth, clip]{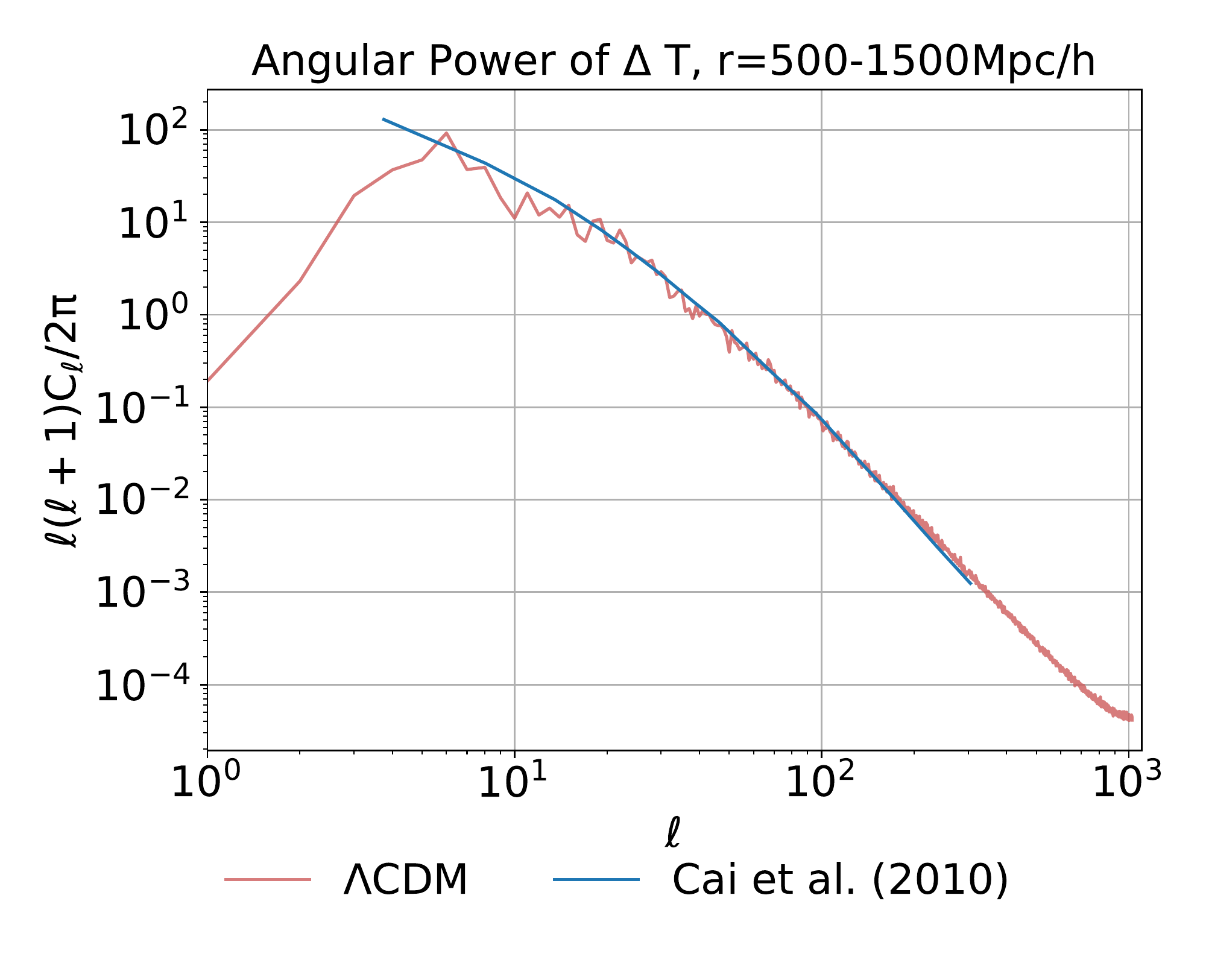}}
   \caption{The ISW induced $\Delta T$ generated using our simulation.The upper panel shows the full-sky temperature fluctuation map, for which the ray-tracing is done between the radial distance [500, 1500] Mpc/h. While in the lower panel we show its corresponding power spectrum. The blue line shows the prediction from \citet{2010MNRAS.407..201C} based on the linear theory, and the red curve is our result. They are well consistent.}
    \label{fig:pk-isw}
\end{figure}

\section{Predicting the ISW effects with a N-body simulation}
\label{sec:simu-data}
Although implementing the LDP analysis is straightforward at the data side, it is highly non-trivial at the theory side due to complicated relation between LDPs and the underlying density/potential field. This is further complicated by various selection effect in observations. Therefore we will use a N-body simulation to generate  ISW maps. The same simulation is used to generate mock galaxy catalogues and LDPs, under given observational conditions. They are then used to predict the ISW-LDP correlation signal. 

\subsection{N-body Simulation}
The ISW signal is mostly contributed from the  large scale mode, so here we use the $1200$Mpc$/h$ N-body simulation from the CosmicGrowth simulation series \citep{2019SCPMA..6219511J}. It contains $3072^3$ simulation particles, and adopts the flat $\Lambda CDM$ cosmology, with $\Omega_c=0.223$, $\Omega_b=0.0445$, $\Omega_\Lambda=0.732$, $\sigma_8=0.83$, h=0.71, and $n_s=0.968$.  Halos are identified with FoF group finder, and subhalos are identified with HBT\citep{2012MNRAS.427.1651H}.

\subsection{Construction of Full Sky ISW Map}
\label{sec:full-sky-isw}
As shown in Eq.\ref{dT2}, the ISW induced $\Delta T$ is an integration of $\dot{\Phi}$ along the line-of-sight. Making use of the Poisson equation\footnote{\begin{equation}
\Phi(\vec{k},t)=-\frac{3}{2}\left(\frac{H_0}{k}\right)^2\Omega_m\frac{\delta(\vec{k},t)}{a},
\end{equation}}, we obtain $\dot{\Phi}$ in Fourier space:
\begin{equation}
\label{Phidot}
\dot{\Phi}(\vec{k},t)=\frac{3}{2}\left(\frac{H_0}{k}\right)^2\Omega_m\left[\frac{\dot{a}}{a^2}\delta(\vec{k},t)-\frac{\dot{\delta}(\vec{k},t)}{a}\right].
\end{equation}
Here $\rho(t)$ is the matter density, $\overline{\rho}(t)$ the mean density, $\delta$ the over-density ($\delta\equiv(\rho-\overline{\rho})/\overline{\rho}$), $H_0$ the current Hubble parameter and $\Omega_m$ the present value of matter density parameter. In the linear regime, $\dot{\delta}(\vec{k},t)=\dot{D}(t)\delta(\vec{k},z=0)$, where $D(t)$ is the linear growth factor. For our purpose, it is sufficient to neglect the nonlinear evolution (Rees-Sciama effect). Therefore $\dot{\Phi}(\vec{k},t) \propto k^{-2}(1-\beta(t))H\delta(\vec{k},t)/a$, where $\beta(t)\equiv dlnD(t)/dlna$.

In simulation, we construct the $\dot{\Phi}$ field in the following way. Firstly, we assign dark matter particles into 3D grids under Cartesian coordinate, and construct the density field $\delta(\vec{x})$. During this process, we use a grid of $512^3$ cells for our simulation box. Then we perform the Fast Fourier Transform on the density field to compute its Fourier form $\delta(\vec{k})$. It is then used to yield the $\dot{\Phi}(\vec{k},t)$ field in Fourier space. At last, we perform the inverse Fourier transform to obtain $\dot{\Phi}(x)$ in real space. Above procedures are repeated for eight output snapshots at redshift 0, 0.058, 0.151, 0.253, 0.364, 0.485, 0.616 and 0.76.  To avoid the discontinuities of $\Phi$ at boundaries, we assume periodic boundary conditions when using our simulation to construct  $\dot{\Phi}(x)$ on a cube whose size length is larger than $1200$ $Mpc/h$. 

Next, we choose the center of spliced cube as the location of observer, and generate angular evenly distributed rays from it using HEALPix with $N_{side} = 512$ resolution. For each ray, we accumulate the temperature fluctuations along its line of sight using Eq.(\ref{dT2}) by taking fixed discrete steps. Note that, values of $\dot{\Phi}$ for the same grid generated from different snapshots are different. So for each step, according to its position on the ray we find out the snapshot at that lookback time, and assign it the value of $\dot{\Phi}$ of its nearest grid. In this way, we are able to construct full sky maps of the ISW effect using the density field.  

In this paper we construct the ISW induced $\Delta T({\widehat{n}})$ with a maximum redshift 0.7. However, there will be two problems arising from the periodic boundary conditions assumed in the above when constructing the $\Delta T({\widehat{n}})$ map. The first problem is that light rays will pass through the same structure every certain distance, leading to larger fluctuations along these directions. This distance is the shortest for light rays along the main axis of the simulation box, equaling to 1200 Mpc/h. The second problem is that the same structure is seen in multi-directions. As pointed out in \cite{2010MNRAS.407..201C} that the first issue can be solved by generating maps with the radial depth less than the simulation boxsize. The second issue does not matter as long as the angular scales of our analysis are less than the angular size subtended by the simulation box.

In the top panel of Fig.\ref{fig:pk-isw} we show the predicted $\Delta T$ map generated by integrating along the line-of-sight of the observer for the range $500<r_c<1500$ Mpc/h, and in the bottom panel we show its power spectrum ( red solid line). The blue solid line is the linear theoretical prediction from \cite{2010MNRAS.407..201C}, which has assumed a very similar cosmology to us. These two lines are found in good consistency. 

\subsection{Generating LDPs with Mock Galaxies}
\begin{figure}
    \centering
    \subfigure{
     \includegraphics[width=1\linewidth, clip]{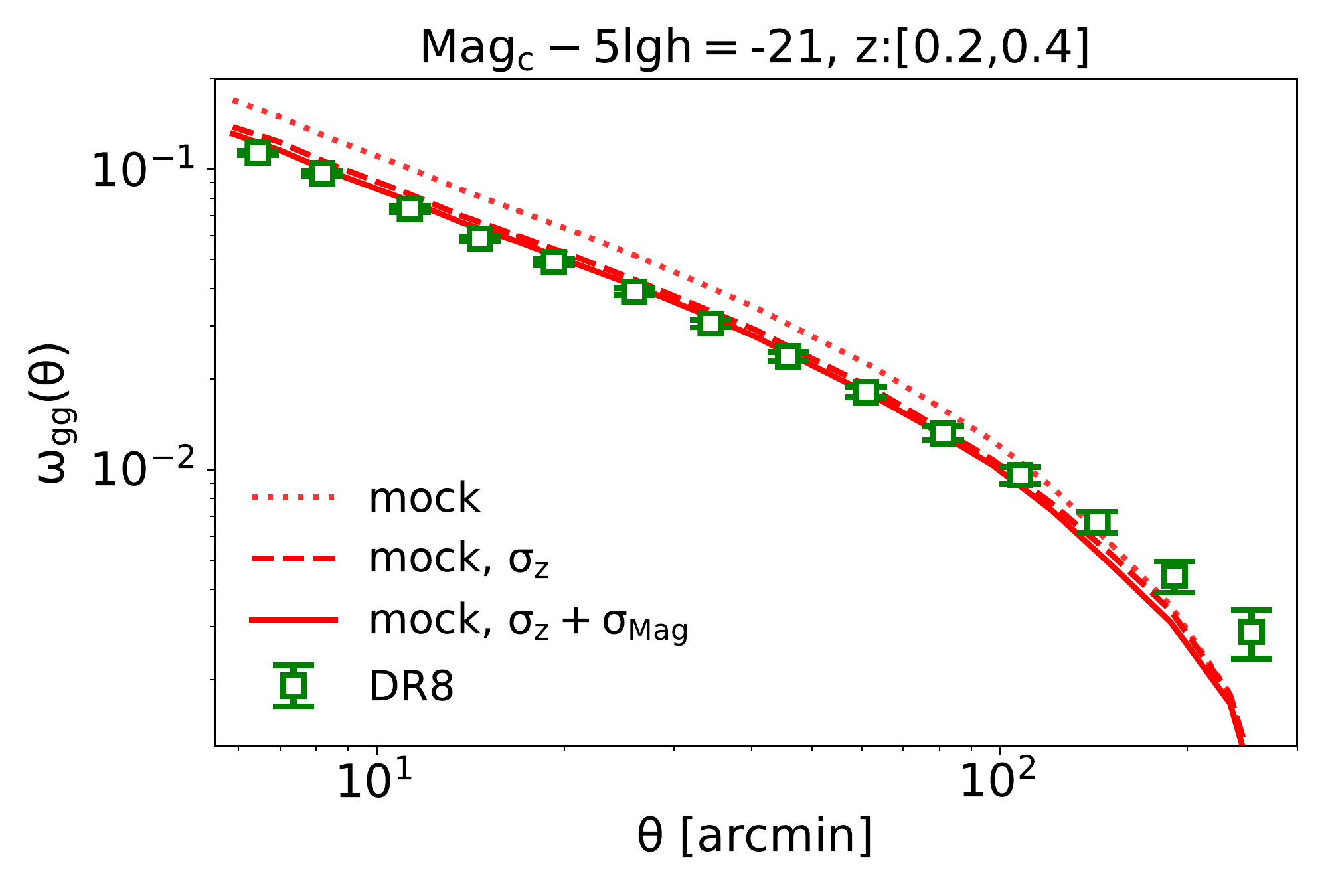}}
   \caption{The auto correlation for galaxies brighter than -21 at $z_m=0.3$. The green data points show the measurement in observation, with jackknife error bars. The red solid line is the result for mock galaxies when both $\sigma_{Mag}$ and $\sigma_z$ are introduced. The red dashed line is the result for mock galaxies when only $\sigma_z$ is introduced. The red dotted line is the result when neither $\sigma_{Mag}$ nor $\sigma_z$ is considered.}
    \label{fig:auto}
\end{figure}
Similar to the operation in \cite{2019ApJ...874....7D}, we draw correspondence between galaxies and halos/subhalos by matching the galaxy-subhalo abundance (SHAM). With absolute magnitudes of galaxies computed in \S\ref{sec:obs}, we measure the luminosity functions for three redshift bins: [0.01,0.2], [0.2,0.4] and [0.4,0.6].  Then for each used snapshot, we compare the number of halos/subhalos with mass at the accretion time greater than $M$ to the number of galaxies with luminosity less than L at that redshift.   
Subhalos from different snapshots are used to fulfill the corresponding radial distance slices. And we adopt the mask of DR8 in simulation in order to ensure the same angular selection of galaxies as in observation. 

To better mimic the real situation, we also add scatters to the redshifts and luminosities of mock galaxies. Firstly, with the probability distribution function of redshift dispersion P($\sigma_{z}$, z) measured in observation\footnote{Considering our galaxy sample, for $z_m=$ 0.1, 0.3 and 0.5, we only use galaxies whose absolute magnitudes are respectively lower than -20, -20.5 and -21 to get P($\sigma_{z}$, z).}, we randomly generate $\sigma_z$ for each galaxy. Then for the purposes of this paper we take the assumption that $p(z_{photo}|z)$ follows a Gaussian function with a zero mean and a scatter $\sigma_z$. We randomly move the positions of galaxies in redshift space and update the absolute magnitudes of mock galaxies according to $z-z_{photo}$. Thirdly, we introduce a constant scatter $\sigma_{Mag}=0.375$ dex to the absolute magnitude to mimic the  galaxy-halo/subhalo relation\citep{2008ApJ...676..248Y}. After introducing these uncertainties, we redo the SHAM to ensure the same luminosity function of mock galaxies as in observation. In this way, we generate the mock photo-z catalogue for $z<0.6$ in simulation. We don't consider for the redshift bin [0.6,0.8] as the generation of its volume-limited mock galaxy sample may suffer from the incompleteness of galaxies at higher redshifts when $\sigma_z$ is introduced. 

To validate our galaxy mocks, we compare the angular distribution of our mock galaxies to observation in Fig.\ref{fig:auto} for $z_m=0.3$. For this purpose, the two-point correlation-function at angular separation $\theta$ is calculated in the simple way of: 
\begin{equation}
\omega_{gg}(\theta)\equiv\langle\delta_g(\widehat{n}_1)\delta_g(\widehat{n}_2)\rangle.
\end{equation}
where $\widehat{n}_i\cdot\widehat{n}_j=cos\theta$ and $\delta_g$ is the number over-density of galaxies. $\delta_g$ is obtained as:
 \begin{equation}
\delta_g(\widehat{n})=\frac{n_g(\widehat{n})-\overline{n}_g}{\overline{n}_g},
\end{equation}
where  $n_g$  is the number density of galaxies in the HEALPix cell with $N_{side}=512$ resolution. 
  The absolute magnitudes of these galaxies are less than -21, and their redshifts are within the slice [0.2, 0.4]. It shows that the $\omega_{gg}$ of mock galaxies is consistent with it in observation. Adding magnitude uncertainty will slightly suppress the correlation function. Adding redshift uncertainty  influences the correlation function more significantly. We also find that the brighter the galaxy, the larger the impact. In \S\ref{sec:conclusion} we will discuss its influence on our LDP-ISW signal measurement.
  
Notice that the auto-correlation function estimator above is by no means optimal and by no means bias-free, comparing to the standard Landy-Szalay estimator. It is only used for the purpose of comparing the mock and data.  Since we use the same estimator for both the mock and the data,  the consistency in $\omega_{gg}$ shows that our galaxy mocks well represent the observed galaxy distribution. 
Since this paper dose not  focus on the auto-correlation analysis, this comparison is sufficient for current purpose. 

\begin{figure}
     \centering
    \subfigure{
     \includegraphics[width=1\linewidth, clip]{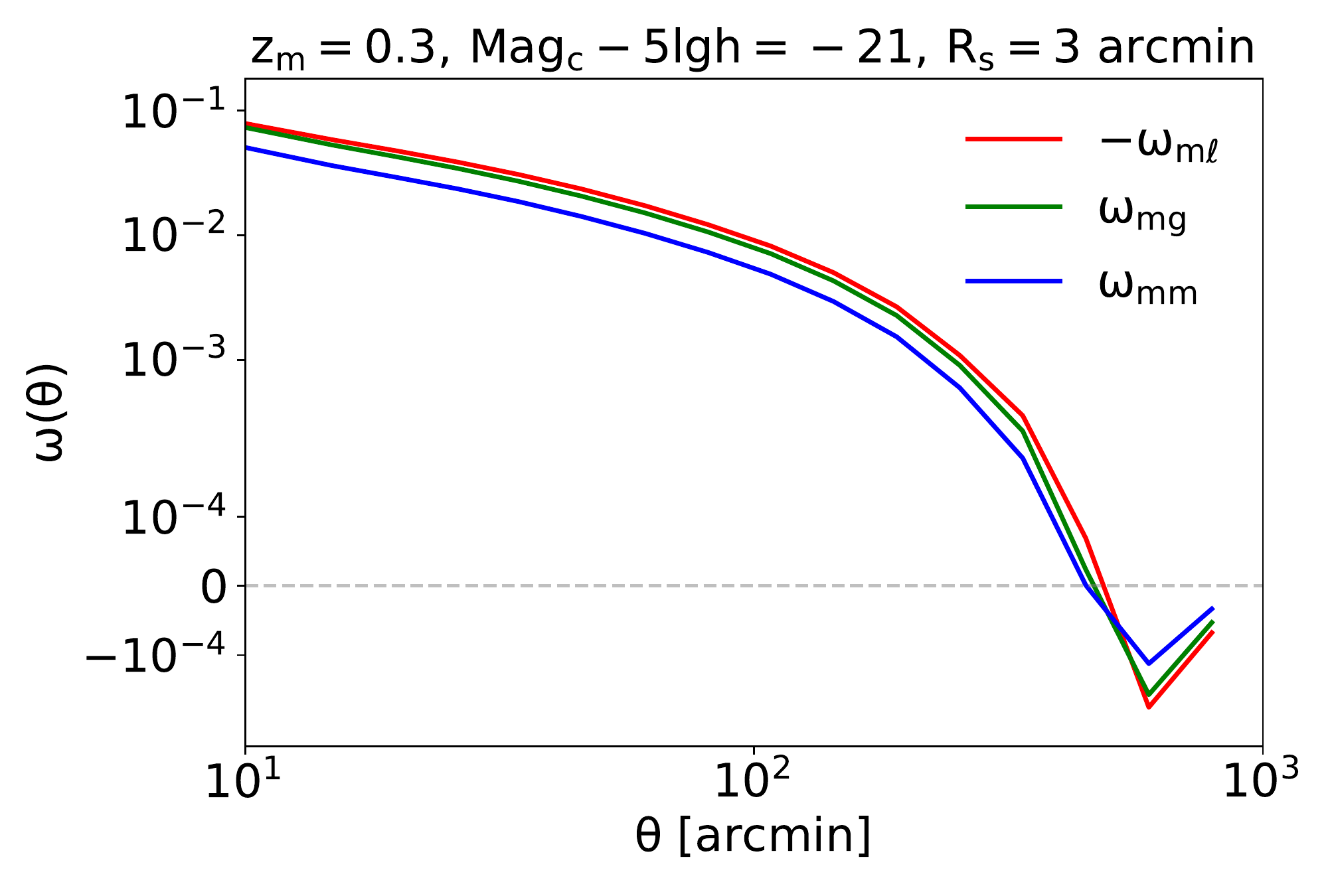}}
   \caption{The cross-correlation between the LDP-matter (red line), galaxy-matter (green line) and matter-matter correlation functions,  in our mock.  The similarity in shapes of the three correaltions and the significance of the $\omega_{ml}$ signal show that the LDP field is indeed tighly correlated with the matter distribution. The negative sign in $\omega_{ml}$ shows that the LDP field indeed probes under-dense regions.}
    \label{fig:gg-bias}
\end{figure}

Then we repeat the operations described in \S\ref{sec:obs} to generate LDPs in simulation. Fig.\ref{fig:gg-bias} shows the cross correlations of LDP overdensity $\delta_l$ with the matter density $\delta_m$. The tight correlation confirms our expectation that $\delta_l$ is indeed a good tracer of LSS. Furthermore, the cross correlation-function has the opposite sign to the matter auto-correlation function or the matter-galaxy cross-correlation function. This confirms that $\delta_l$ is a good tracer of low-density regions of the universe. 

\section{ISW-LDP cross correlation measurements}
\label{sec:result}

With the data analysis tools and simulation tools presented in previous sections, we now proceed to the LDP-ISW cross correlation measurements and their theoretical interpretation.

\begin{figure}
    \centering
    \subfigure{
     \includegraphics[width=1\linewidth, clip]{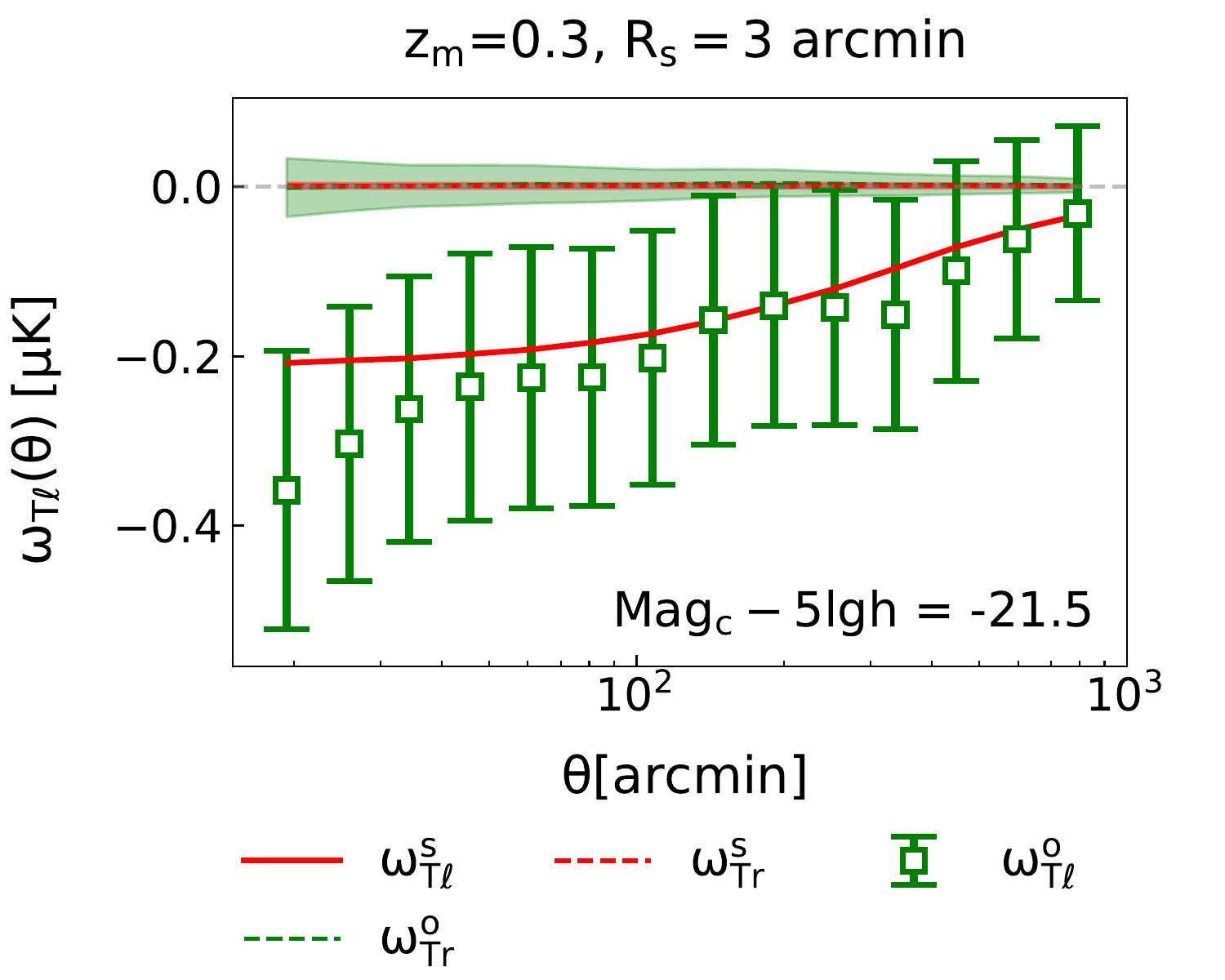}}
   \caption{ The LDP-CMB cross correlation measured for $z_m=0.3$. LDPs are generated with galaxies whose absolute magnitudes are less than -21.5 and $R_s=\,3^{'}$.  The green square points show $\omega^o_{T\ell}$ from observation and the red solid line is the prediction from simulation. The green and red dashed lines show results of the null test ($\omega_{Tr}$) respectively for observation and simulation. They are calculated by correlating randomly disturbed $\delta_l$ with $\Delta T$. Error bars shown for $\omega^o_{Tl}$ are estimated with the CMB rotating strategy. The green (red) shadow area shows one $\sigma$ range of $\omega^o_{Tr}$ ($\omega^s_{Tr}$).}
    \label{fig:cross}
\end{figure}

\begin{figure*}
    \centering
     \subfigure{
     \includegraphics[width=1\linewidth, clip]{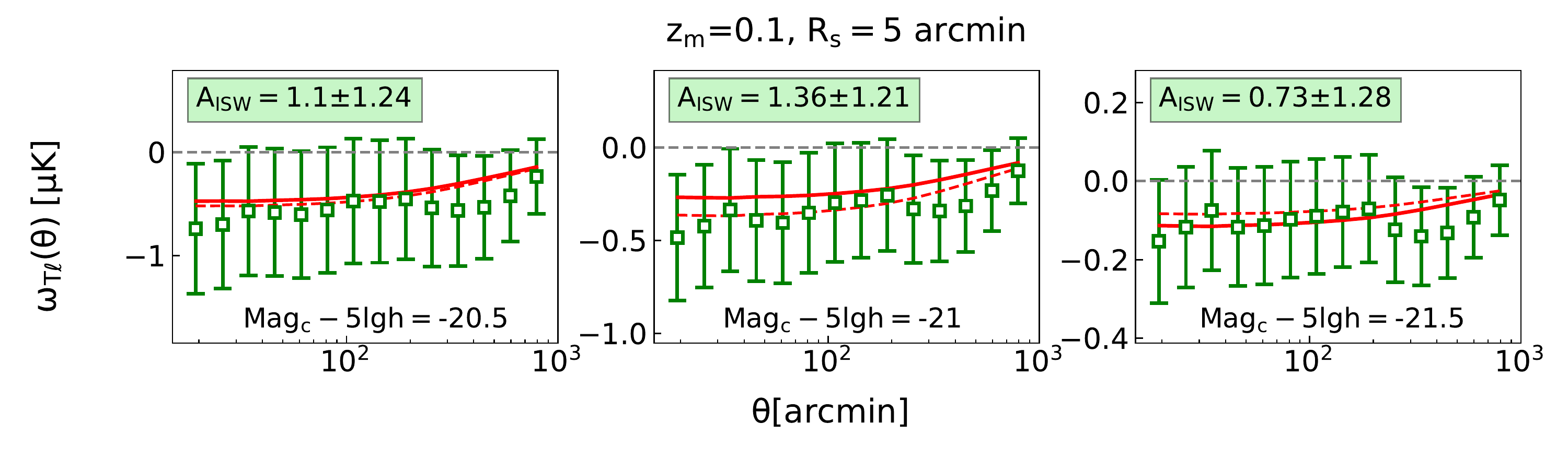}}
    \subfigure{
     \includegraphics[width=1\linewidth, clip]{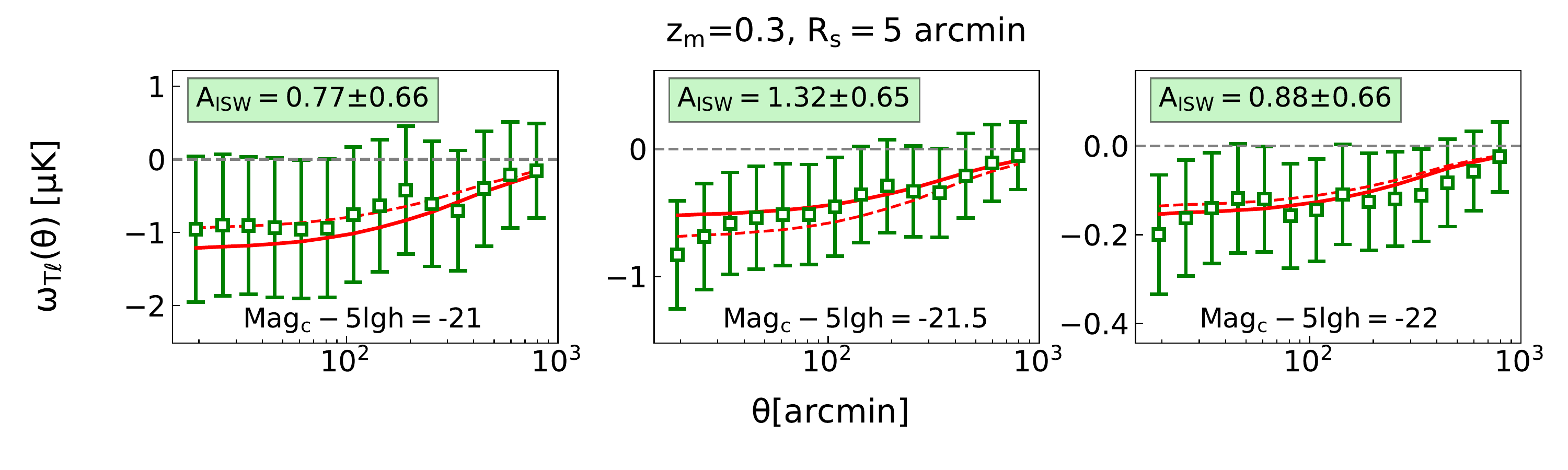}}
     \subfigure{
     \includegraphics[width=0.68\linewidth, clip]{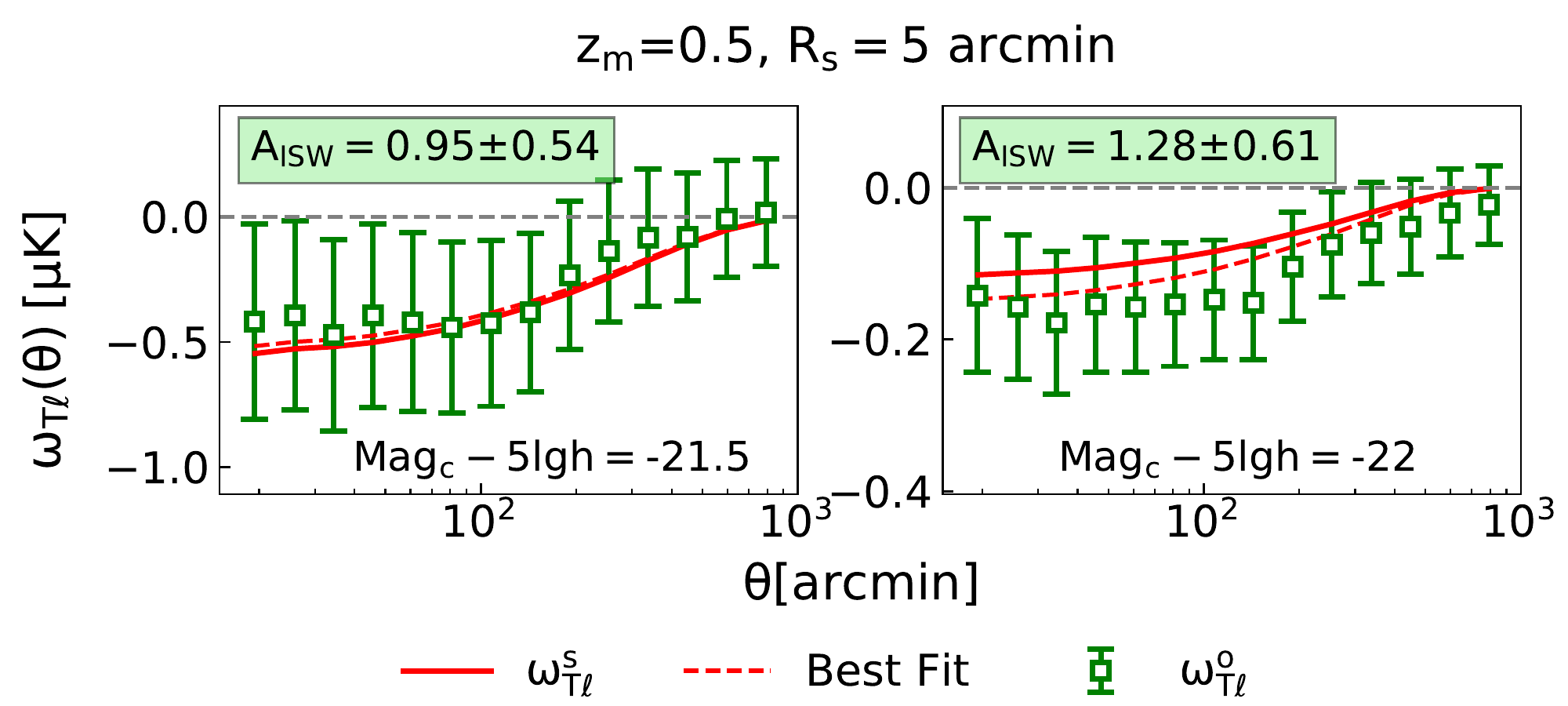}}
   \caption{The LDP-CMB cross correlation measured for three redshifts: $z_m=$ 0.1, 0.3 and 0.5. $R_s$ is set to 5 arcmin. For redshift $z_m=0.1$, the left, middle, and right columns are for $Mag_c-5lgh=$ -21.5, -21, -20.5 respectively. For $z_m=0.3$, the three columns are for $Mag_c-5lgh=$ -22, -21.5 and -21. And for $z_m=0.5$, the two columns are for $Mag_c-5lgh=$ -22 and -21.5. }
    \label{fig:cross-all5}
\end{figure*}

\subsection{The cross-correlation measurements}
\label{sec:result-cross}
We adopt a simple estimator for the angular correlation-function $\omega_{T\ell}(\theta)$ between the CMB temperatue map and the LDP over-density $\delta_l$ map, 
\begin{equation}
\label{eq-tl}
\omega_{T\ell}(\theta)=\langle\delta_T(\widehat{n}_1)\delta_\ell(\widehat{n}_2)\rangle,
\end{equation}
and we evaluate it with TreeCorr package\citep{2015ascl.soft08007J}. $\delta_T(\widehat{n})=T(\widehat{n}) - T_0$ is the temperature fluctuation, and $\delta_\ell$ the LDP overdensity. With the data  in both observation and simulation, we have $4$ cross-correlation measurements. $\omega^o_{T\ell}$ is calculated with DR8 galaxy catalogue. $\omega^o_{Tr}$ is obtained by randomly shuffling $\delta_l$ in different cells and correlated with $\Delta$T. And this operation is repeated for 100 times to obtain the average value and variance. Here the superscript ``o'' denotes observation. $\omega^o_{Tr}$ is useful for the null-test and, if non-zero, should be subtracted from $\omega^o_{T\ell}$ to correct the mean impact of various selection effects such as the survey geometry, mask and mean CMB fluctuations. Namely, the finally estimated ISW-LDP cross correlation is
\begin{eqnarray}
\label{eqn:TTr}
\hat{\omega}_{T\ell}(\theta)=\omega^o_{T\ell}(\theta)-\omega^o_{Tr}(\theta)\ .
\end{eqnarray}
Correspondingly, $\omega^s_{Tl}$ and $\omega^s_{Tr}$ are calculated with mock catalogue in simulation.

\begin{figure*}
    \centering
     \subfigure{
     \includegraphics[width=1\linewidth, clip]{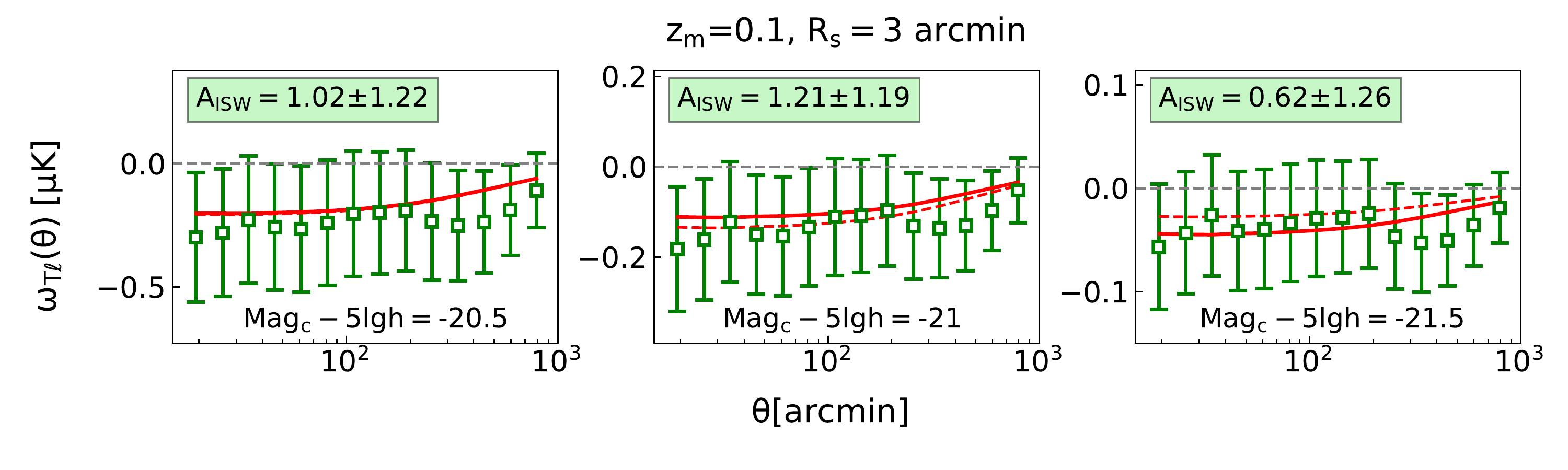}}
    \subfigure{
     \includegraphics[width=1\linewidth, clip]{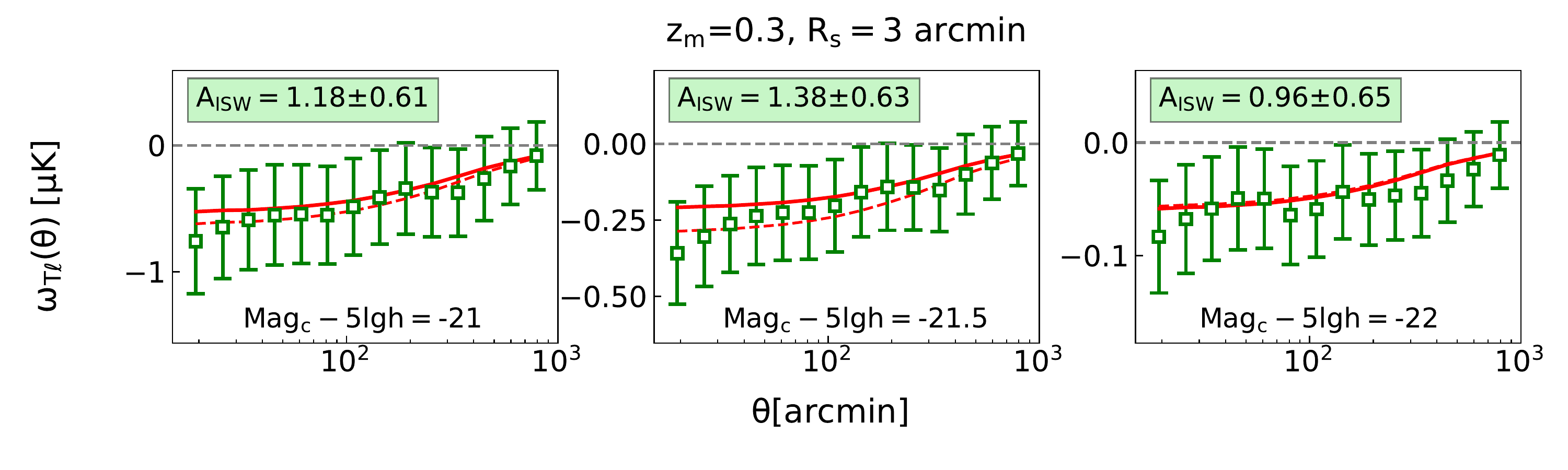}}
     \subfigure{
     \includegraphics[width=0.68\linewidth, clip]{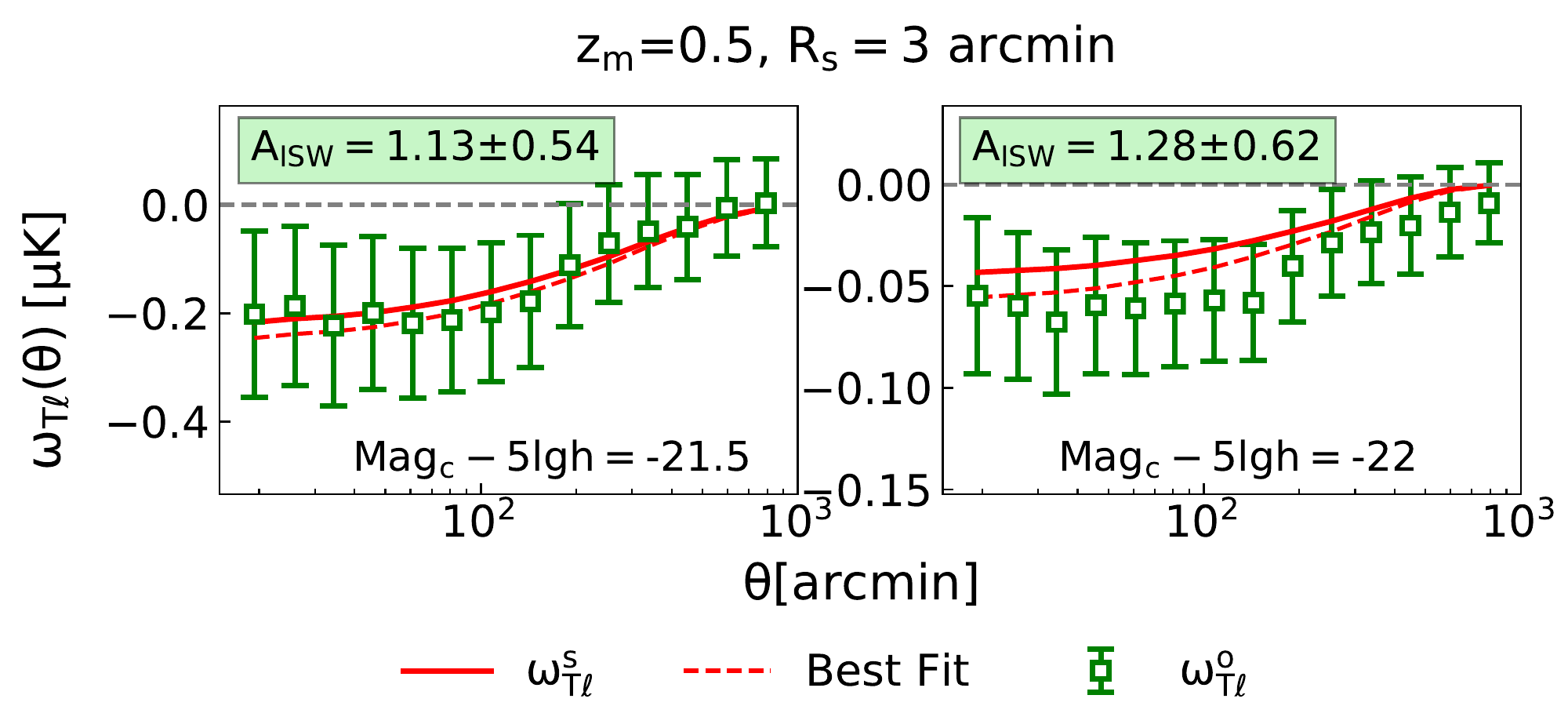}}
   \caption{Similar to Fig.\ref{fig:cross-all5}, but with $R_s$ shrinked to 3 arcmin.}
    \label{fig:cross-all3}
\end{figure*}

Fig.\ref{fig:cross} shows one of such measurements, in which the LDPs are generated with parameters $Mag_c-5lgh=-21.5$, $R_s=$3 arcmin and $z_m=0.3$. The first finding is that $\omega^o_{Tr}\simeq 0$. The scatter is $\sim 2\times 10^{-3}\mu K$, about 10 percent of $\omega^o_{Tl}$.  Therefore it does not matter wether $\omega^o_{Tr}$ is subtracted or not.  $\omega^s_{Tr}$ is also consistent with zero ($1\times 10^{-3}\mu K$), with much smaller $\sigma$ as no CMB components/foregrounds are considered in our simulation. The second finding is that, the measured cross correlation is in good agreement with the theoretical prediction $\omega^s_{Tl}$.\footnote{When given the prediction of $\omega^s_{T\ell}$ from simulation, there is one issue that needs further attention. Considering that photo-z errors are added to our mock galaxies, there will be galaxies from other redshifts entering the redshift bin $[z_m-0.1, z_m+0.1]$ used for generating LDPs. So the ISW induced temperature fluctuations from the neighboring redshifts will associate with $\delta_l(z_m)$. In this case, it is better to generate $\delta_T(\widehat{n})$ by integrating temperature fluctuations within a broader redshift range. However, to avoid the enlarged $\Delta T(\widehat{n})$ problem caused by the repeated structures discussed in \S\ref{sec:full-sky-isw}, we choose to generate the full-sky map using the $\dot{\phi}$ field within $[z_m-0.2, z_m+0.2]$. The redshift depth 0.4 is considered both from the fact that $\overline{\sigma}_z\sim0.1$ and the boxsize of simulation.} 
We repeat the above calculations for three redshift bins: $z_m=$0.1, 0.3, 0.5, and for different choices of critical magnitude.  Fig.\ref{fig:cross-all5} \& \ref{fig:cross-all3} show the results of $R_s=5^{'}$ and $3^{'}$ respectively.  In general we find good agreement between observation and theory/simulation. 

\begin{figure}
     \centering
    \subfigure{
    \includegraphics[width=1\linewidth, clip]{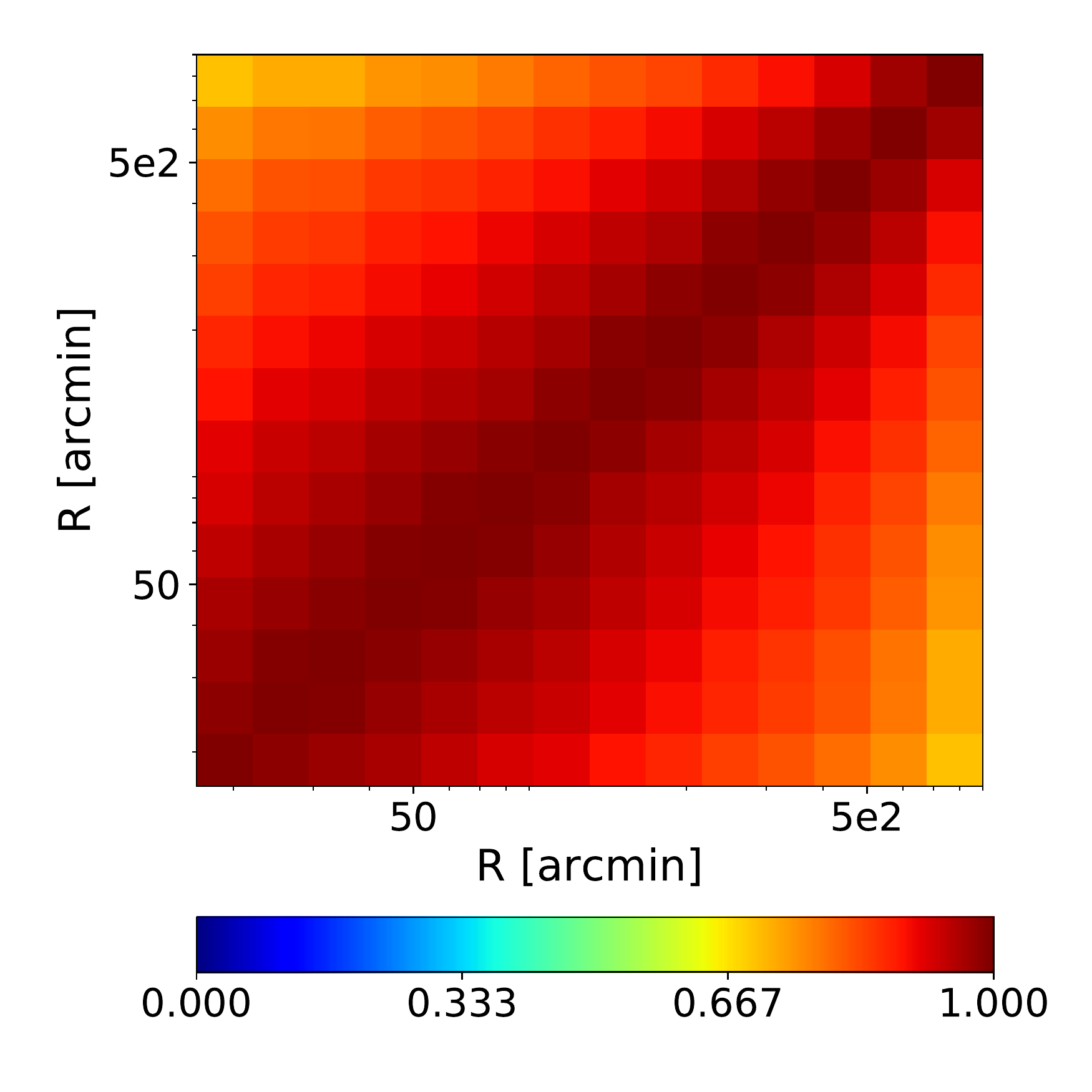}}
   \caption{The covariance matrix for $\omega^o_{Tl}$. It is calculated with the $\it Rot.$ strategy for parameters $z_m=0.3$, $Mag_c-5lgh=-21.5$ and $R_s=5'$. Further discussions and more results are presented in the appendix. }
    \label{fig:cv}
\end{figure}
\subsection{The covariance estimation and the detection significance}
\label{sec:cov}
Error bars shown in Fig.\ref{fig:cross} are directly estimated from the CMB map by rotating it around different axis with angle $\Delta\Phi$. This is based on the consideration that the ISW induced temperature fluctuation is much smaller than the fluctuation from the original CMB. The rotation strategy has been proposed in \cite{10.1111/j.1365-2966.2009.16054.x}, in which the WMAP maps have been rotated around the galactic pole to identify galactic contamination. However, any choice of axis is feasible, since one does not expect correlations for a large rotation angle $\Delta\Phi$. So scatters from independent rotations should reflect the intrinsic variance in the measurements, including the instrumental effect. To perform independent measurements, a minimum rotation  $\Delta\Phi=30^\circ$ is required, as suggested in \cite{2012MNRAS.426.2581G}. Therefore, each rotation axis leaves 11 independent samples of cross-correlation measurement. We choose 18 positions on CMB map as the rotation axis \footnote{ We make the following sets of ($\theta$, $\phi$) as rotation axis: $\theta= [90\degree, 120\degree, 150\degree, 180\degree, 210\degree, 240\degree, 270\degree, 300\degree, 330\degree]$ , $\phi=$[30\degree, 60\degree]. }, with angular distances between them no less than $30^\circ$. In this way, we get 198 independent measurements of $\omega_{T\ell}$. The covariance matrix is estimated by 
\begin{eqnarray}
\label{eq-cov}
C_{ij}&=&\frac{1}{N}\sum_{n=1}^{N}[(\omega_{T\ell,n}(\theta_i)-\overline{\omega}_{T\ell}(\theta_i))\nonumber \\
&&\times (\omega_{T\ell,n}(\theta_j)-\overline{\omega}_{T\ell}(\theta_j))].
\end{eqnarray}
We show the normalized covariance matrix $C_{ij}/\sqrt{C_{ii}C_{jj}}$ for $\omega^o_{T\ell}$ in the lower panel of Fig.\ref{fig:cv}. It indicates strong correlations between different angular scales, which is a common feature for correlation-function measurement.   $\sigma_{\omega}=C^{1/2}_{ii}$ is the error bar shown in Fig.\ref{fig:cross}, for each angular bin $\theta_i$.

\begin{table*}
    \footnotesize
    \centering
    \caption{S/N of $\omega^o_{T\ell(g)}$.}
    \label{table:sn}
\begin{tabular}{ccccccccccccc}
\hline
&&&& \multicolumn{4}{c}{S/N}&Max($S/N_t$) \\ 
\hline
COV &$R_s$[arcmin]& {\diagbox[innerleftsep=-0.5cm,innerrightsep=0pt,innerwidth=3.6cm]{$z_m$}{$Mag_c-5lgh$}} && -20.5 &-21 &-21.5 &-22 &\\
\hline
\multirow{3}{*}{{\it Rot.}} &\multirow{3}{*}{3} &0.1 && 1 & 1.1 & 0.9 & -   &\multirow{3}{*}{3.2}  \\
&&0.3 && - & 1.9 & 2.2 & 1.5      \\
&&0.5 && - & - & 2.1 & 2.1          \\ 
\hline
\multirow{3}{*}{{\it Rot.}} &\multirow{3}{*}{5} &0.1 && 1 & 1.2 & 0.9 & -   &\multirow{3}{*}{3.2} \\
&&0.3 &&- & 1.2 & 2.1 & 1.4       \\
&&0.5 && - & - &1.8  &2.1           \\
\hline

\multirow{3}{*}{{\it Jack.}} &\multirow{3}{*}{3} &0.1 &&1.5  &1.4 & 1.2  &-   &\multirow{3}{*}{3.7}\\
&&0.3 &&- & 2.1 & 2.5 & 2         \\
&&0.5 &&- &- & 2.3 &2.3            \\
\hline
\multirow{3}{*}{{\it Jack.}} &\multirow{3}{*}{5} &0.1 &&1.4 &1.5 &1.2 &-   &\multirow{3}{*}{3.6}\\
&&0.3 &&- &1.3 &2.3 & 1.8        \\
&&0.5 &&- &- &1.8 & 2.3           \\
\hline
\multirow{3}{*}{{\it Rot.}}&\multirow{3}{*}{galaxy} &0.1 && 0.9 & 1 & 0.9 &-&\multirow{3}{*}{3.4}\\
&&0.3 && - & 2.4 & 2.3 & 1.5     \\
&&0.5 && - & - & 2.2 & 2.1         \\
\hline
\multicolumn{9}{l}{`{\it Rot.}' represents for using the CMB-rotation technique to estimate the covariance matrix.}&\\
\multicolumn{8}{l}{`{\it Jack.}' represents for using the Jackknife technique.}&\\
\multicolumn{8}{l}{The last three rows are for CMB-galaxy correlation measurement.}&\\
\end{tabular}
\end{table*}

\begin{table*}
    \footnotesize
    \centering
    \caption{The bestfit $A_{ISW}$, the associated $1$-$\sigma$ uncertainty,  and $\chi^2_{min}$, combining all three redshifts.}
    \label{table:aisw}
\begin{tabular}{ccccccccccc}
\hline
&&& \multicolumn{4}{c}{$\langle A_{ISW}\rangle$}&\multicolumn{4}{c}{$\rm{\chi^2_{min}}$}\\  
\hline
COV &$R_s$[arcmin]&  &$Max(Mag_c-5lgh)$ &-21.5 &$Min(Mag_c-5lgh)$ &&&$Max(Mag_c-5lgh)$ &-21.5 &$M in(Mag_c-5lgh)$\\
\hline
\multirow{1}{*}{{\it Rot.}} &3.        & & 1.14$\pm$0.38 & 1.18$\pm$0.39 &1.07$\pm$0.42    &&& 0.36 & 0.89 & 1.1  \\
\multirow{1}{*}{{\it Rot.}} &5         & & 0.9$\pm$0.4     & 1.07$\pm$0.4   &1.06$\pm$0.42    &&& 0.29 & 0.79 & 1   \\
\multirow{1}{*}{{\it Rot.}} &galaxy & &1.25$\pm$0.38 & 1.21$\pm$0.38  &1.07$\pm$0.42    &&& 0.69 & 1 & 1.2   \\
\hline
\multicolumn{10}{l}{For $z_m=0.1,0.3,0.5$, $Max(Mag_c-5lgh)$ equals to -20.5, -21,-21.5 respectively.}&\\
\multicolumn{10}{l}{For $z_m=0.1,0.3,0.5$, $Min(Mag_c-5lgh)$ equals to -21.5, -22,-22 respectively.}&\\
\end{tabular}
\end{table*}

For each galaxy/LDP sample,  the total S/N of the observational signal can be calculated as
\begin{equation}
\label{eq:sn}
\frac{S}{N}=\left[\sum_{i,j}\,\omega^o_{T\ell}(\theta_i)C^{-1}_{ij}\omega^o_{T\ell}(\theta_j)\right]^{1/2}\ .
\end{equation}
The results are shown in Table \ref{table:sn}. 
LDP sampes of different photo-z bins are uncorrelated, so the total S/N of the ISW measurement of all three photo-z bins ($\alpha=1,2,3$) is 
\begin{equation}
(S/N)_t = \sqrt{\sum_{\alpha=1}^3(S/N)_\alpha^2}.
\end{equation}
Depending on the choice of LDP samples, the total S/N of three photo-z bins varies. 
For example, when $R_s=5^{'}$,  $(S/N)_t=2.9$ for a universal magnitude cut $Mag_c=-21.5$. For other combinations of magnitude cut, $2.3\leq (S/N)_t\leq 3.2$.  For $R_s=3^{'}$,  $2.7\leq (S/N)_t\leq 3.2$. Therefore we have achieved a measurement of the ISW effect induced by low-density regions of the universe, at a significane of $3.2\sigma$.

This $3.2\sigma$ detection is comparable to the $3.4\sigma$ detection directly using galaxies (Table \ref{table:sn}).  This is an interesting point to address and to further investigate. First, this  LDP-ISW cross-correlation is not equivalent to directly using galaxies, since the LDP overdensity $\delta_l$-galaxy overdensity $\delta_g$ relation is nonlinear and non-local. For this reason, it is not subject to the $\sim 7\sigma$ upper bound of S/N in galaxy-ISW cross correlation measurement, for ideal CMB/galaxy surveys \citep{2004PhRvD..70h3536A}. Second, given that the weighting that we adopt to convert LDPs to $\delta_l$ is not optimal, the S/N should further improve if we find and apply the optimal weighting.  This is an issue for future investigation. 

We caution that the estimated S/N depends on the estimation of covariance matrix and its inversion . We will discuss another estimation of covariance matrix using the Jackknife resampling method, and  the inversion of noisy covariance matrix with the SVD method. The results shown in the main text use the covariance matrix estimated by rotating CMB (hereafter $\it Rot.$), and the inversion by keeping the first two eigenmodes of the covariance matrix. Table \ref{table:sn} shows the S/N of other choices.

\begin{figure*}
     \centering
    \subfigure{
     \includegraphics[width=1\linewidth, clip]{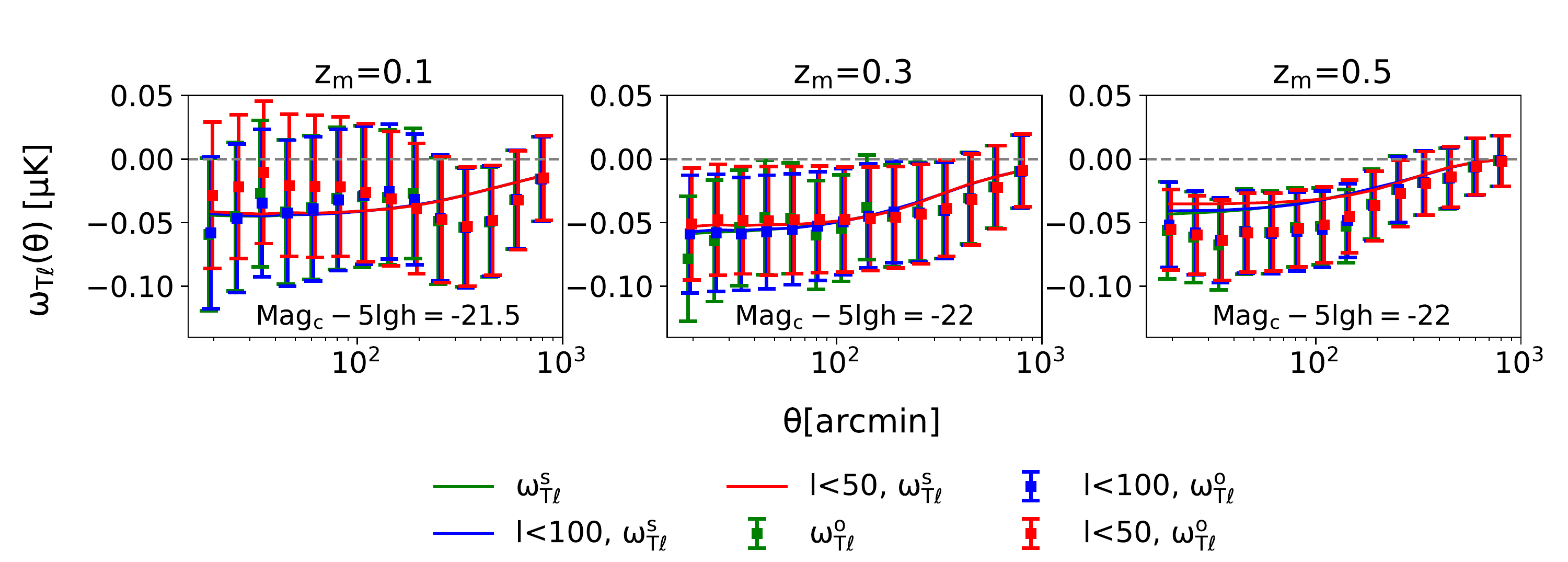}}
   \caption{Testing the origin of the detected cross-correlation,  by a tophat cut in multipole $\ell$ space ($\ell_{cut}=50$, $100$, and no cut).   The cut $\ell_{cut}=100$ has essentially no impact on $\omega_{T\ell}$, showing that the cross-correlation signal mainly arises from $\ell<100$.  In contrast, the cut $\ell_{cut}=50$ causes visible difference in $\omega_{T\ell}$, showing that contribution from $50<\ell<100$ is non-negligible. These dependences are consistent with the ISW origin.}
    \label{fig:lcut}
\end{figure*}

\subsection{Comparision with theoretical prediction}
Clearly, Fig. \ref{fig:cross-all5} \& \ref{fig:cross-all3} show that the measurements are in good agreement with the prediction of the concordance $\Lambda CDM$ cosmology. To further quantify the agreement, we choose the model of fitting as 
\begin{equation}
\omega^o_{T\ell}=A_{ISW}\omega^s_{T\ell}\ .
\end{equation}
Here $A_{ISW}$ is the amplitude to fit and the standard $\Lambda$CDM cosmology has $A_{ISW}=1$.  We minimize 
\begin{eqnarray}
\chi^2&=&\delta \omega({\theta_i})C^{-1}_{ij}\delta \omega({\theta_j}), \nonumber\\
\delta_\omega(\theta)&\equiv& \omega^o_{T\ell}-\omega^s_{T\ell}\ .
\end{eqnarray}
to obtain the bestfit $A_{ISW}$. The bestfit $A_{ISW}$ and its $1\sigma$ uncertainty is shown in Fig.\ref{fig:cross-all5} \& \ref{fig:cross-all3}.  We find that constrained $A_{ISW}$ of all LDP samples are  consistent with unity, within 1$\sigma$ statistical uncertainty. Therefore the concordance $\Lambda$CDM indeed describes the data excellently. 
 
We further combine all three redshifts to constrain $A_{ISW}$. The constraint depends on which sample is used for each redshift. If using the largest galaxy sample at each redshift,  we obtain
\begin{eqnarray}
A_{ISW}(R_s=3')&=&1.14\pm0.38, \nonumber\\
A_{ISW}(R_s=5')&=&0.9\pm0.4.
\end{eqnarray}
In Table \ref{table:aisw} we show $A_{ISW}$ estimated from various combinations, all consistent with unity. It also shows the minimum chi-square $\chi^2_{min}$ corresponding to are all smaller than $1.2$, demonstrating excellent agreement with $\Lambda$CDM. 
 In contrast to previous constraints of $A_{ISW}\gg 1$\citep{2010MNRAS.401..547H,2010ApJ...724...12I,2012JCAP...06..042N,2014A&A...572C...2I,2014ApJ...786..110C,2015MNRAS.446.1321H,2015MNRAS.452.1295K,2017MNRAS.465.4166K,2018MNRAS.475.1777K,2019MNRAS.484.5267K}, we find no tension with the concordance $\Lambda$CDM. Furthermore, this agreement holds for different thresholds of $R_s$ (and therefore under-density).


\subsection{Further consistency checks}
The detected signal arises from the low-density regions of the universe, but free of many selection effects of void identification. Nevertheless, given the known difficulties in ISW measurements, we carry out three more tests in order to further validate the measurements.

\begin{itemize}
\item First is to check the origin of the cross-correlation. The ISW effect is expected to arise from large scale and is insensitive to small scale CMB modes. Fig. \ref{fig:lcut} shows the measured LDP-CMB cross correlation with a tophat cut $\ell_{cut}=50$, $100$ in multipole $\ell$ space. We find that the results with $\ell_{cut}=100$ are almost identical to the ones without a cut. Therefore the measured cross-correlation indeed comes from the large angular scale, as expected. Furthermore, the cut $\ell_{cut}=50$ results in minor but visible loss of the cross correlation signal, in particular for the lowest redshift bin $0.01<z<0.2$. This is again expected if the signal arises from the ISW effect. The $k$ corresponding to a given $\ell$ is roughly $k\sim \ell/r(z)$. For the same $\ell$, lower redshift means larger scale (smaller $k$) and therefore larger loss of signal. 

\item We also check the measured galaxy density-ISW cross correlation, using the same galaxy samples and the same analysis pipeline. The overall S/N reaches $3.4$ (Table \ref{table:sn}). And the results are also in good agreement with the $\Lambda$CDM prediction, $A_{ISW}=1.21\pm0.38$ for $Mag_c-5lgh=-21.5$. Since the galaxy overdensity field and the LDP overdensity field are defined very differently, both agreements with $\Lambda$CDM further validate our measurements.  
\item  Besides the Planck SMICA map, we have other CMB maps to analyze. We have tried the Planck 100 GHz map and the V-band WMAP map. After subtracting the foregrounds, the results are consistent.
\end{itemize}

Therefore we believe the robustness of our LDP-ISW detection. The excellent consistency with $\Lambda$CDM then implies hidden systematics in some of the void-ISW cross correlation measurements. Therefore this LDP method is highly complementary to existing methods to cross-check and improve the ISW measurement.

\section{Summary \& Discussion}
\label{sec:conclusion}

We have designed a novel method of ISW measurement, by cross-correlating LDPs (low-density-positions, \citet{2019ApJ...874....7D}) and CMB. We then apply it to the DESI imaging survey DR8 galaxy catalogue of BASS + MZLS + DeCALS + DES, and Planck SMICA map. We achieve a $3.2\sigma$ detection of the ISW effect (Table 2), one of the most significant among existing measurements. Furthermore, the detected signal is fully consistent with the concordance $\Lambda$CDM prediction ($A_{ISW}=1$) for  all the galaxy samples that we investigated and the adopted LDP definitions (Table 3), with the bestfit $A_{ISW}$ consistent with $A_{ISW}=1$ and $\chi^2\in (0.3,1.1)$. For example, for $Mag_c-5lgh=-21.5$ and $R_s=3^{'}$, we  find $A_{ISW}=1.18\pm0.39$, with $\chi^2_{\min}=0.89$. For $R_s=5^{'}$, $A_{ISW}=1.07\pm0.4$ and $\chi^2_{\min}=0.79$.

The achieved S/N ($3.2\sigma$) is already competitive to that with galaxy-ISW cross correlation ($3.4\sigma$ that we have measured), and there exists room for further improvement. Together with the excellent agreement with the concordance cosmology, we have demonstrated the applicability of the LDP method to measure the  ISW effect, for the first time. Our measurement provides an independent check to existing tensions between void ISW and $\Lambda$CDM, and between void ISW and galaxy ISW. Since our LDP ISW measurement has no tension with $\Lambda$CDM prediction and galaxy ISW measurement, we suggest hidden systematics in void ISW measurements. 

The measurement can be used to constrain dark energy, in particular given a flat geometry. There are potentially other applications. For example, galaxy overdensity and LDP overdensity probe regions of the universe with statistically different matter density/gravitational potential. So the combination of LDP ISW and galaxy ISW may probe beyond $\Lambda$CDM physics, such as clustered dark energy and screening phenomena in modified gravity models. 

Although the measurement S/N is already high among existing ISW detections, there are still possible improvements, related to the LDP definition and the LDP overdensity definition. 
\begin{itemize}
\item LDP definition. LDPs depend on the galaxy sample, which is in turn determined by the redshift range and radius threshold, magnitude cut, and other galaxy properties. We have only tried a few configurations and the obtained S/N is unlikely optimal. 
\item LDP overdensity $\delta_l$ definition. The underlying matter density $\delta_m$ at LDPs is statistically negative. But the exact value varies with LDPs. For example, $\delta_m$ at LDPs surrounded by LDPs should be on average more negative than LDPs surrounded by non-LDPs. Intuitively speaking, $\delta_m$ should decrease monotonically with increasing $d_{min}$, the distance of a given LDP to the nearest galaxy. The overdensity defintion (Eq. \ref{eqn:LDPoverdensity}) reflects this expectation. It is indeed tightly (negatively) correlated with the underlying matter density field, as verified with our simulation (Fig. \ref{fig:gg-bias}). However, although it has enabled a $3.2\sigma$ detection of the ISW effect, it is still an open question on whether it is the optimal choice.  For example, for the adopted definition of $\delta_l$, its value equals to $Max(\delta_l)$ in void regions of size $\gg 6.87^{'}$. So it downweights the contribution from large voids. The optimal $\delta_l$ definition must take it into account.   From the viewpoint of ISW measurement, the optimal definition of $\delta_l$ should result in a cross-correlation coefficient  with the gravitational potential field as close to unity as possible. 
\item The way to populate galaxies.  In this work we use the SHAM method to populate galaxies in simulation by allowing a scatter $\sigma_{Mag}$ between galaxy luminosity and halo/subhalo mass and a scatter $\sigma_z$ between the true galaxy redshift and photometrical redshift. Adding $\sigma_{Mag}$ will decrease the nominal absolute magnitude on average, as more fainter galaxies are mistaken for bright ones statistically. While adding $\sigma_z$ will increase the nominal absolute magnitude on average, as more brighter galaxies at higher redshifts are mistaken as fainter galaxies at lower redshifts.  So these two uncertainties would lead to the so-called Eddington bias\citep{1913MNRAS..73..359E}.  For example,  when $z_m=0.3$ and $Mag_c-5lgh=-21$, the number of galaxies increases by $5\%$, while the galaxy sample becomes 1.2 times larger for $Mag_c-5lgh=-22$. Our solution  is to redo the SHAM after introducing these uncertainties. Otherwise, the distribution of galaxies with magnitude is changed.   Although this is still a rough model to populate galaxies, the comparison of correlation-function in Fig.\ref{fig:auto} show the rationality of our operation.
\item The estimator of cross correlation function.  We adopt a simple cross correlation function estimator. More optimal estimator requires input of galaxy selection function. DESI imaging survey galaxy catalogue still contains various imaging systematics, not fully captured by the random catalogue that we use \citep{2020MNRAS.tmp.1781K}. The resulting non-uniform selection function biases the galaxy auto correlation measurement. The problem for the cross correlation that we perform in this paper is much less severe, since the galaxy selection function is uncorrelated with CMB and ISW. Furthermore, to a good approximation, it is uncorrelated to residual foreground in the Planck CMB map. Nevertheless, non-uniform galaxy selection function amplifies statistical error in the cross correlation measurement. Future work needs to suppress such error (and diagnose potential systematics) by improving the cross correlation estimator with the aid of random catalogue.  
\end{itemize}
Besides these measurement issues, accurate determination of the covariance matrix and robust estimation of the S/N and $A_{ISW}$ is also important. In the appendix, we have presented our treatments on the covariance matrix and its inverse. We plan to further investigate these issues, with the aid of numerical simulations and mock catalogues. 

There are other possibilities to further explore. In principle the cross-correlation in real space is identical to that in Fourier (spherical harmonic) space. But in reality, due to the scale cut, mask and noise/foreground distribution, the two can differ. In this paper we only work on the real space, and leave the analysis in Fourier space elsewhere. For the theory/simulation side, we have used SHAM to populate galaxies into N-body simulation. This exercise turns out to be successful.  Nevertheless, there may still room of improvement for higher S/N and better theoretical prediction. 

\section*{Acknowledgements}

We thank Yipeng Jing for providing us the N-body simulation. We also thank Jian Yao and Ji Yao for useful discussions.  This work is supported by the National Key Basic Research and Development Program of China (No.15ZR1446700, 2018YFA0404504,19ZR1466800), the NSFC grants (11621303,11653003,11673016,11833005,11890692,11773048), the 111 project (No. B20019). 

\vspace{6pt}

\bibliographystyle{mnras}
\bibliography{ISW}
\appendix

\begin{figure*}
     \centering
     \subfigure{
     \includegraphics[width=1\linewidth, clip]{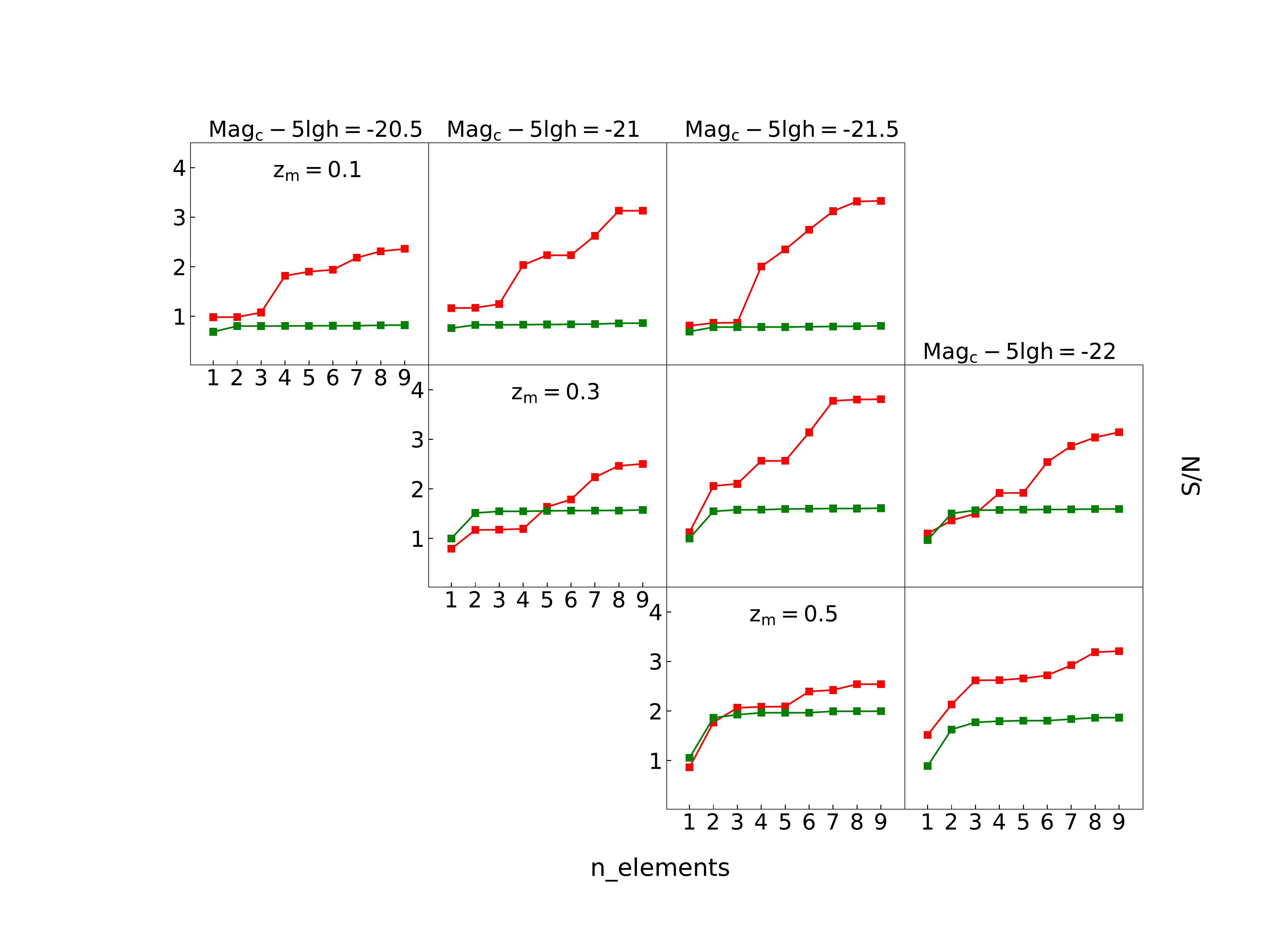}}
   \caption{ The estimated S/N by keeping the first $n_{elements}$ eigenmodes in the SVD procedure. The red line is for $\omega^o_{T\ell}$, and the green line is for $\omega^s_{T\ell}$. The critical radius $R_s$ equals to 5'. This tests suggest to keep the first 2-3 eigenmodes. }
    \label{fig:svd-snr}
\end{figure*}

\section{Performing Singular-Value Decomposition to Covariance Matrix}
\label{appendix:svd}
In \S\ref{sec:cov} we find that S/N of the observational signal $\omega^o_{T\ell}$ is overestimated compared to the one calculated with the theoretical line $\omega^s_{T\ell}$, using the same covariance matrix. 
Here we perform the Singular-Value Decomposition (SVD) on $C_{ij}$ for further analysis:
\begin{equation}
	\mathbf{C}=\mathbf{U} \Lambda \mathbf{U^{T}}\ .
\end{equation}
Since the covariance matrix  $C$ is symmetric and real, all the eigenvalues $\lambda_i$ are real. Since it is positive definite, all $\lambda_i>0$.  $\Lambda=$diag$(\lambda_i)$, and $\lambda_i$ is the $i$-$th$ eigenvalue ($\lambda_1\geq \lambda_2\cdots$). $U$ is a rotation orthogonal matrix ($\bf{U^TU = I}$), with the $i$-th eigenvectors of $C$ as its $i$-$th$ column vectors.  Noise in $C$ contaminates these eigenvectors. The ones with smaller eigenvalues have larger errors, which are further amplified in $C^{-1}$.  Together with noise in $\omega$, this causes error in the estimated S/N. The SVD (or pseudo-inverse) then modifies $C^{-1}$ to 
\begin{equation}
\bf{C}={\bf U}\ {\rm diag}(\lambda^{-1}_1,\lambda^{-1}_2,\cdots, \lambda_M^{-1},0,\cdots)\  {\bf U}^{T}\ .
\end{equation}
Namely, it disregards the $i>M$ eigenvectors  in the inverse of ${\bf \Lambda}$ to stablize the inverse and to reduce the impact of noise in $C$. The task now is to determine $M$. 

\begin{figure*}
     \centering
     \subfigure{
     \includegraphics[width=0.8\linewidth, clip]{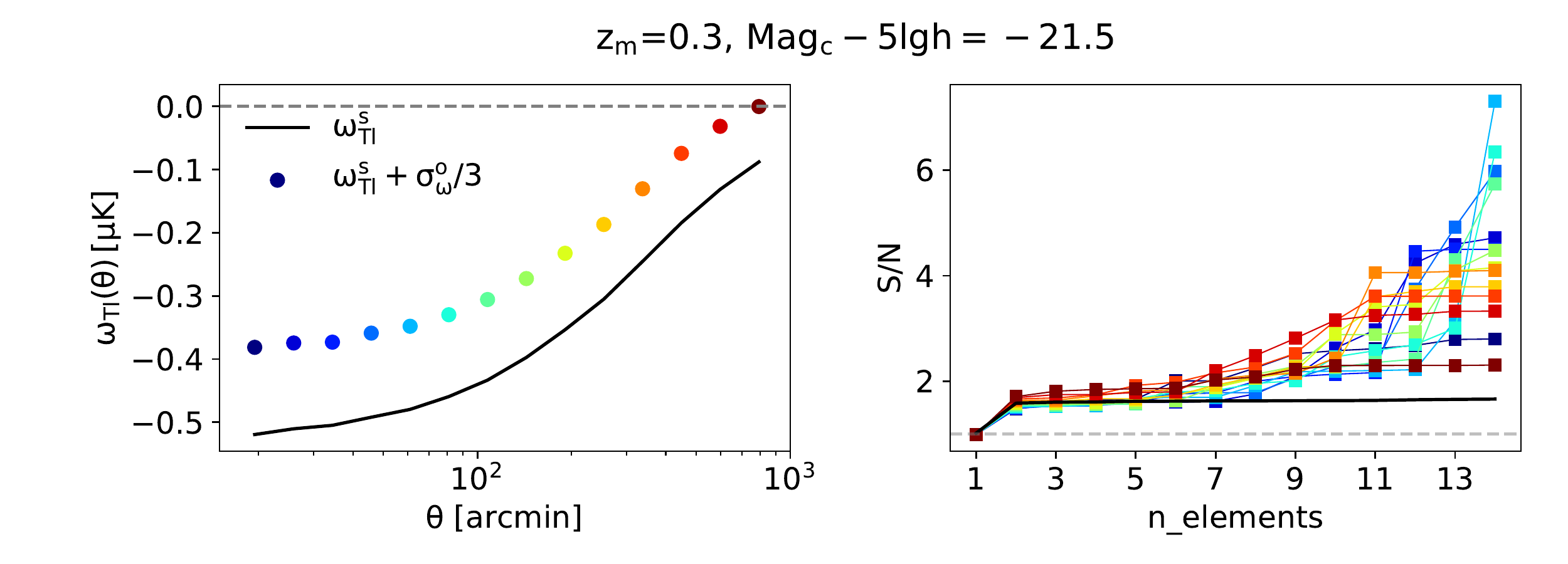}}
   \caption{ The stability of S/N (right panel) against a small bump artificially added to $\omega^s_{Tl}$ (left panel). This test rules out the possibility of keeping the eigenmodes $i\gtrsim 5$, and supports the option of keeping the first 2-3 eigenmodes.}
    \label{fig:snr-bump}
\end{figure*}

${\bf C}$ in our case is a $N\times N$ ($N=14$) matrix. The one shown in  Fig.\ref{fig:cv} has $\lambda_i=12.74, 0.82, 0.21, 0.09, 0.05, 0.02,\cdots$. We have checked that the first 2-3 eigenmodes captures major features of ${\bf C}$. But the other eigenmodes cause large variation in ${\bf C}^{-1}$ and S/N, due to the $1/\lambda$ operation in ${\bf C}^{-1}$. We perform two convergence tests to determine $M$. 
\begin{itemize}
\item Fig.\ref{fig:svd-snr} shows the S/N of $\omega^{o(s)}_{T\ell}$ as a function of $M$.  S/N of $\omega^o_{T\ell}$ is sensitive to the $M>2$ eigenmodes, despite the fact that these eigenmodes has insignificant contribution to ${\bf C}$. In contrast, the S/N of  $\omega^s_{T\ell}$ becomes very stable in all cases when $M\geq2$ These results suggest $M=2,3$ as a reasonable choice of SVD and S/N estimation. 
\item  We also test the stability of S/N against artificial bump ($\sigma^o_{\omega}/3$) added to $\omega^s_{T\ell}$, each at a single $\theta$ (Fig.\ref{fig:snr-bump}). Since the added bump is only $1/3$ of the statistical scatter, we expect insignificant change in S/N, otherwise something in ${\bf C}^{-1}$ may be wrong. We find that for $M=2,3$ the S/N is indeed stable. 
\end{itemize}
Therefore $M=2,3$ are reasonble choices for our situation. $M=3$ leads to a larger S/N. To be conservative, we adopt $M=2$ to estimate the S/N. 


\section{Jack-knife Resampling}
\label{appendix:jackknife}
In this section, we estimate error bars using the jackknife resampling technique by dividing the whole sky into fields at a resolution of $N_{side} = 4$ (resolution equals to 879 arcmin). If the number of LDPs within a field is too few either due to the mask effect or the edge effect, we will merge it into one of its neighboring field, whose number of LDPs is the fewest. In this way, the number distribution of LDPs in different fields become more homogeneous. The CMB pixels are also divided into the same fields. Jack-knife subsamples are then generated by removing one field at onetime, with $C_{ij}$ given as:
\begin{eqnarray}
\label{eq-cov}
C_{ij}&=&\frac{N_{J}-1}{N_{J}}\sum_{n=1}^{N_{J}}[(\omega^J_{T\ell,n}(\theta_i)-\overline{\omega^J}_{T\ell,n}(\theta_i))\nonumber \\
&&\times(\omega^J_{T\ell}(\theta_j)-\overline{\omega^J}_{T\ell}(\theta_j))],
\end{eqnarray}
where $\omega^J_{T\ell,n}$ is the cross-correlation measured for the n-th subsample, $\overline{\omega^J}_{T\ell,n}$ is the average of subsamples' measurements, and i and j refer to the i-th and j-th radial bins. 


\begin{figure}
    \centering
    \subfigure{
     \includegraphics[width=1\linewidth, clip]{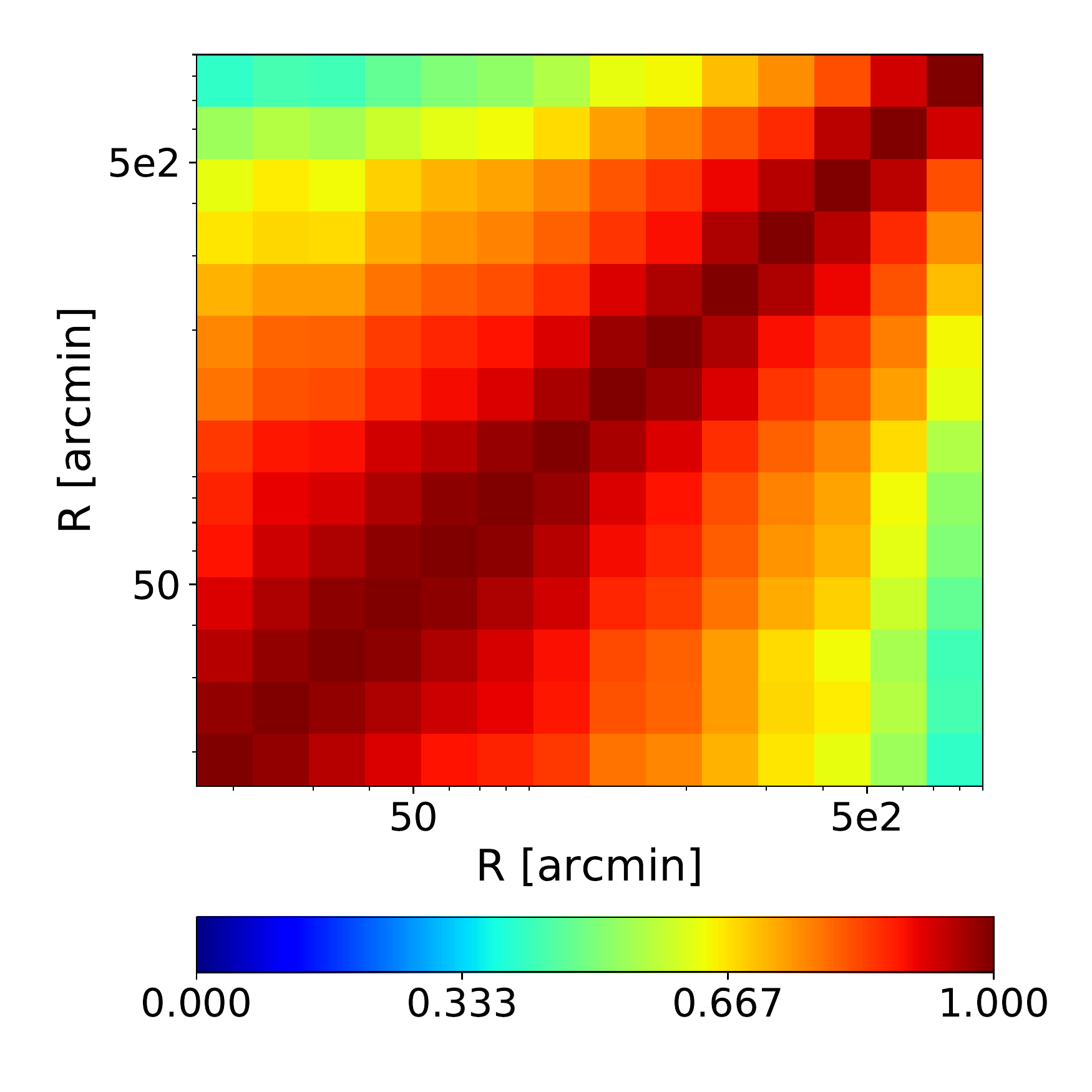}}
   \caption{Similar to Fig.\ref{fig:cv}, but estimated using the jackknife resampling technique.}
    \label{fig:cmb-cov}
\end{figure}

With these covariance matrics, we recalculate S/N. As shown in Table \ref{table:sn}, they are found $10\%$-$20\%$ higher than those estimated with $\it Rot.$ technique, since smaller error bars and lower off-diagonal terms of $C_{ij}$ are found with jack-knife resampling technique. One possible explanation is that jack-knife error would underestimate ISW errors, as pointed out in \cite{2005PhRvD..72d3525P}. To be conservative, we do not adopt these higher S/N.

\section{ISW Signal measured with Galaxies}
\label{appendix:ISW-galaxy}

\begin{figure*}  
    \centering
     \subfigure{
     \includegraphics[width=1\linewidth, clip]{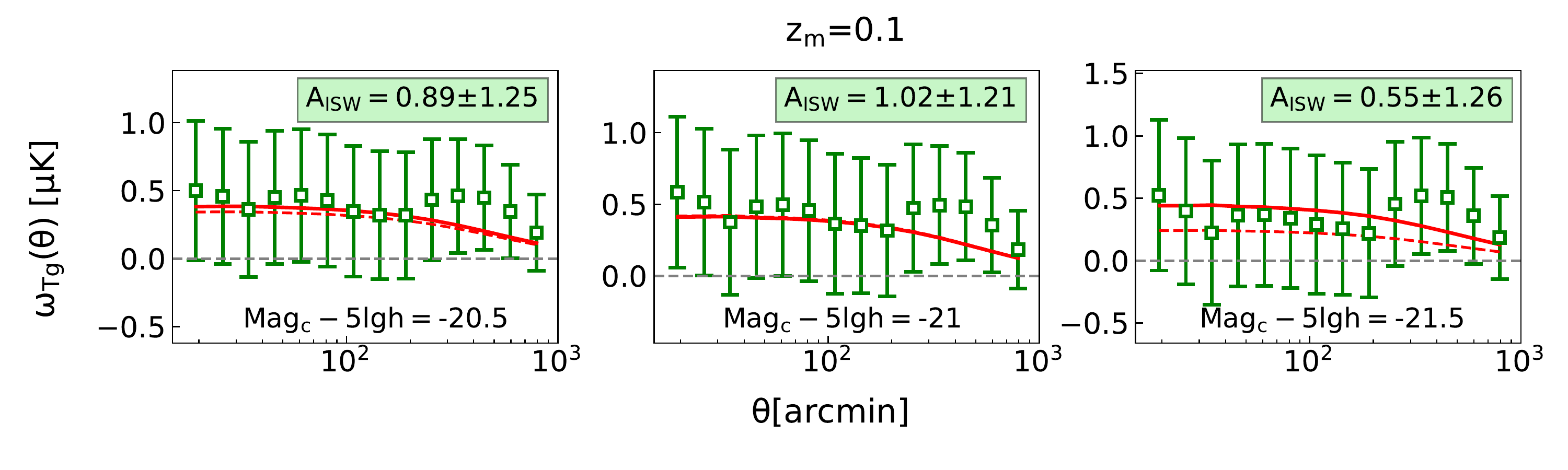}}
    \subfigure{
     \includegraphics[width=1\linewidth, clip]{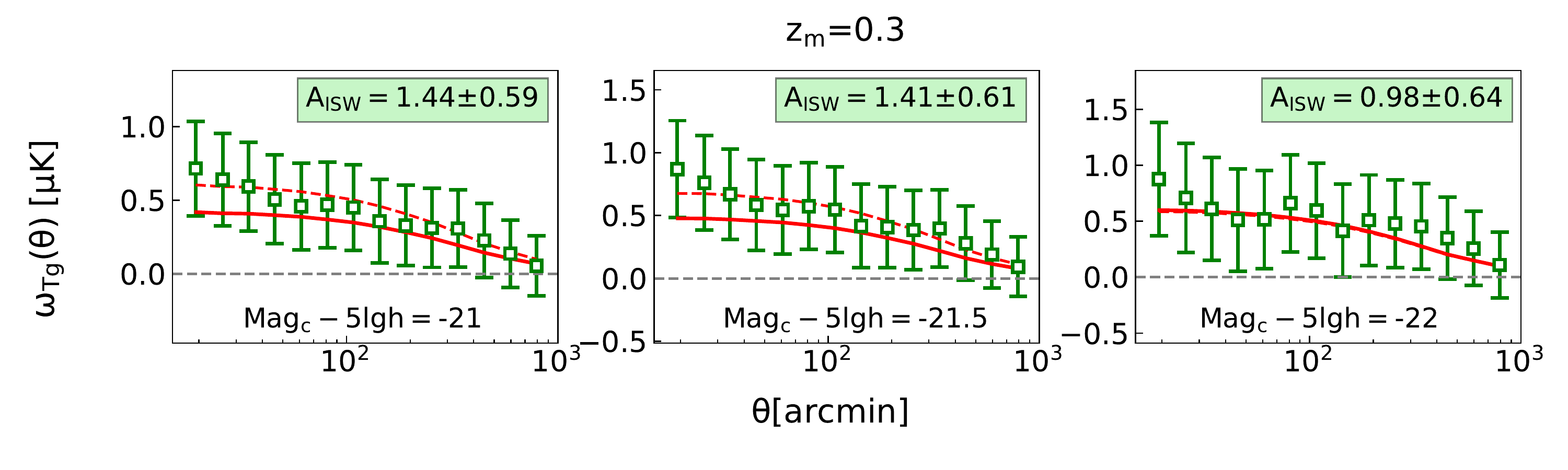}}
     \subfigure{
     \includegraphics[width=0.68\linewidth, clip]{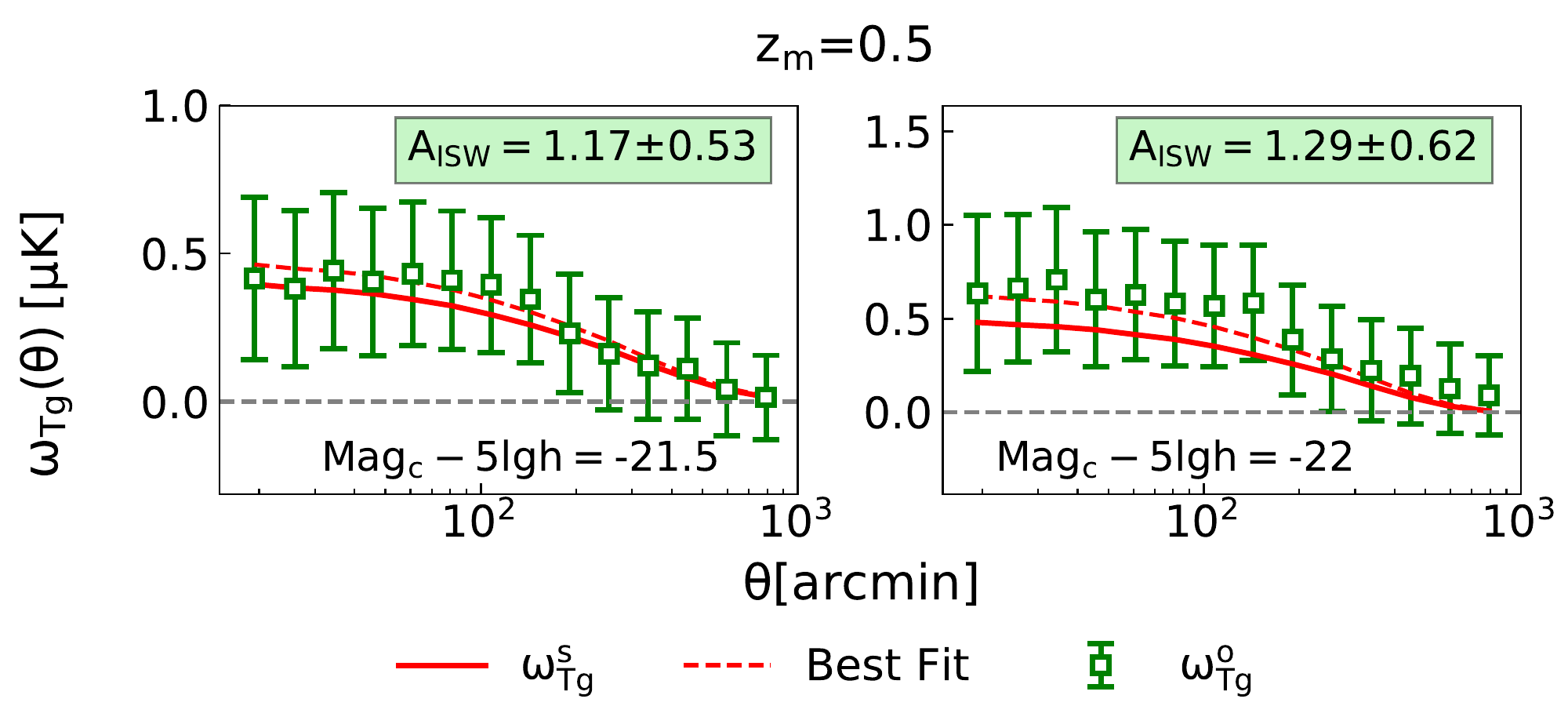}}
   \caption{The cross-correlation between galaxies and temperature fluctuations. The panels are arranged in a similar way with Fig.\ref{fig:cross-all5}.}
    \label{fig:cross-all3-gal}
\end{figure*}

Although the findings of ISW effect detected with galaxies seem to be conclusive, situation is more complicated. In  \cite{10.1111/j.1365-2966.2009.16054.x}, negative signals are measured for $z=0.7$ with luminous red galaxy (LRG) sample, while good consistent is found  between data and the standard ISW model for lower redshifts. This conclusion is further confirmed by \cite{2019arXiv190911095A} with larger galaxy sample. In \cite{2016JCAP...09..003K}, ISW-LRG signals have been found to be higher  than $\Lambda CDM$, and an evolved halo bias has to be introduced to solve the discrepancy. In \cite{2010A&A...513A...3L}, the author concludes that there is no evidence yet of a significant detection of the integrated Sachs-Wolfe (ISW) effect after repeating the analyses in some papers, since field-to-field fluctuations are found to dominate the detected signals. Much of the uncertainty in these studies arises from the Poisson noise in the galaxy distribution. So one of the key to further verify these problems is to increase the survey area to decrease the statistical errors. 

Considering that the current sky coverage of DESI DR8 galaxy catalog approaches 20000 $deg^2$, in this section we revisit the CMB-galaxy correlation. The results are shown in Fig.\ref{fig:cross-all3-gal}, in which the grouping of galaxies is same to Fig.\ref{fig:cross-all5}. The error bars are estimated using $\it Rot.$ technique. One can find that these signals are close in amplitudes, $\sim$ 0.5 $\mu K$ for $\theta<50^{'}$, although they corresponds to different galaxy samples. In general, they are consistent with the prediction curves from simulation. The $A_{ISW}$ measured in each panel is close to 1.


\bsp	
\label{lastpage}
\end{document}